\documentclass[
 preprint,
 superscriptaddress,
 amsmath,amssymb,
 aps,
pre,
 showkeys
]{revtex4-1}
\usepackage{graphicx}
\usepackage{dcolumn}
\usepackage{bm}
\usepackage{amsmath}
\usepackage{hyperref}
\hypersetup{colorlinks=true,linkcolor=black,citecolor=black,urlcolor=black}
\usepackage{color}

\newenvironment{psmallmatrix}
  {\left(\begin{smallmatrix}}
  {\end{smallmatrix}\right)}
\newcommand*\diff{\mathop{}\!\mathrm{d}}

\DeclareMathOperator{\diag}{diag}

\newcommand{\vect}[1]{\bm{#1}}
\newcommand{\matr}[1]{\mathbf{#1}}
\newcommand{\multiindex}[1]{\mathbf{#1}}

\newcommand{\average}[2][]{\left\langle #2\right\rangle_{#1}}

\newcommand{\pfrac}[2][\unskip]{\frac{\partial #1}{\partial #2}}
\newcommand{\ddfrac}[2][\unskip]{\frac{\diff #1}{\diff #2}}

\newcommand{\ZZ}{\mathbb{Z}}
\newcommand{\ii}{\bm{i}}

\newcommand{\mb}[1]{\bm{#1}}

\newcommand{\bq}{\multiindex{q}}
\newcommand{\br}{\multiindex{r}}
\newcommand{\bd}{\multiindex{d}}
\newcommand{\bs}{\mb\sigma}
\newcommand{\bA}{\matr{A}}
\newcommand{\bB}{\matr{B}}
\newcommand{\bC}{\matr{C}}
\newcommand{\bP}{\matr{P}}
\newcommand{\bR}{\matr{R}}

\newcommand{\dtau}{\diff \tau}

\newlength{\templength}

\usepackage[normalem]{ulem}

\begin{document}
\graphicspath{{./Figures/}}


\title{Extending the linear-noise approximation to biochemical systems influenced by intrinsic noise and slow lognormally distributed extrinsic noise}

\author{Emma M. Keizer}
\affiliation{Laboratory of Systems \& Synthetic Biology, Wageningen UR, Wageningen, The Netherlands}
\author{Bj\"{o}rn Bastian}
\affiliation{Institute for Ion Physics and Applied Physics, University of Innsbruck, Innsbruck, Austria}%
\author{Robert W. Smith}
\affiliation{Laboratory of Systems \& Synthetic Biology, Wageningen UR, Wageningen, The Netherlands}
\author{Ramon Grima}
\affiliation{School of Biological Sciences, University of Edinburgh, Edinburgh, United Kingdom}%
\author{Christian Fleck}
\affiliation{Laboratory of Systems \& Synthetic Biology, Wageningen UR, Wageningen, The Netherlands}
\affiliation{Current address: Department of Biosystems Science and Engineering, ETH Zurich, Basel, Switzerland}

\date{\today}

\begin{abstract}
It is well known that the kinetics of an intracellular biochemical network is stochastic. This is due to intrinsic noise arising from the random timing of biochemical reactions in the network as well as due to extrinsic noise stemming from the interaction of unknown molecular components with the network and from the cell's changing environment. While there are many methods to study the effect of intrinsic noise on the system dynamics, few exist to study the influence of both types of noise. Here we show how one can extend the conventional linear-noise approximation to allow for the rapid evaluation of the molecule numbers statistics of a biochemical network influenced by intrinsic noise and by slow lognormally distributed extrinsic noise. The theory is applied to simple models of gene regulatory networks and its validity confirmed by comparison with exact stochastic simulations. In particular we show how extrinsic noise modifies the dependence of the variance of the molecule number fluctuations on the rate constants, the mutual information between input and output signalling molecules and the robustness of feed-forward loop motifs. 
\end{abstract}

\maketitle


\section{\label{sec:intro}Introduction}
Studying the dynamics of biological systems is essential in understanding the design principles underlying biochemical and synthetic networks. However, this task is also challenging given the complexity of interactions between the system's components and its environment. At the molecular level, biological processes possess a certain degree of randomness as chemical reactions are probabilistic events~\cite{Gillespie77}. This stochasticity or \textit{noise} in biological networks has widely varying functional roles, and can be both advantageous and detrimental to cells. Positive effects include phenotypic diversity and the ability to quickly adapt to changing environmental conditions, thus increasing the probability of survival~\cite{Raser05, Rao02}. In contrast, noise can also restrict the ability of a cell to resolve input signals of different strengths and hence reduces the information that can be accessed about the external environment ~\cite{Cheong11}. For these reasons, the network's topology either exploits or attenuates noise.

Due to the high complexity of the intracellular machinery, one often can only study a set of reactions between a certain number of observable molecular components. We call this subset of reactions and components \textit{a biochemical system of interest}. The rest of the intracellular and extracellular reactions, species and environmental cues not accounted for, we call \textit{the dynamic environment}. Noise in the biochemical system's dynamics can then stem from either itself or from the dynamic environment. That originating from itself, i.e. from the inherent discreteness of the molecules participating in the biochemical system is called \textit{intrinsic noise} \cite{VanKampen}; this type of noise increases with decreasing average molecule numbers and hence is particularly relevant to intracellular dynamics due to the low copy number of genes, messenger RNA (mRNA) and some proteins in a single cell ~\cite{Elowitz02}. The noise stemming from the dynamic environment is termed \textit{extrinsic noise} and this affects the biochemical system of interest via modulation of its rate constants. Several studies have shown that a consideration of both types of noise is crucial to understanding gene expression and biochemical dynamics in both prokaryotic and eukaryotic systems~\cite{Elowitz02, Raser04, Fleck07}.

 The well-mixed stochastic description of point-particle biochemical systems is given by the chemical master equation (CME) \cite{VanKampen}. However, this equation rarely can be solved exactly for gene regulatory networks with feedback interactions and when it can, it invariably is for the case of zero extrinsic noise \cite{Grima12,KumarPlatiniKulkarni_PRL2014}. There are two distinct ways to proceed: (i) exact stochastic simulation or (ii) approximate analytic techniques, which we discuss in this order next.

The stochastic simulation algorithm (SSA), as formulated by Gillespie \cite{Gillespie77}, provides trajectories which are consistent with the CME provided the rate constants are time-independent, i.e, it can only describe intrinsic noise since extrinsic noise manifests as noise in the rate constants. The SSA is noteworthy because it is exact in the limit of an infinite number of samples. Modifications to the SSA that take into account both types of noise have been devised by Shahrezaei \textit{et al.}~\cite{Shahrezaei08}, Anderson~\cite{Anderson_JCP2007} and by Voliotis \textit{et al.}~\cite{Voliotis15}. The latter is the most computationally efficient algorithm of the three and also the only one not suffering from numerical integration error. However, the disadvantage of these methods is that a large number of simulations may be required to obtain statistically significant results~\cite{Rao02}.

 Alternatively, various approximation schemes have been derived to obtain analytical expressions for statistical moments and the marginal distributions of molecular numbers (for a recent review see~\cite{SchnoerrSanguinettiGrima_JPA2017}). Scott \textit{et al.}~\cite{Scott06} and Toni \textit{et al.}~\cite{Toni13} develop approximate methods based on the linear-noise approximation~\cite{VanKampen} which allow the calculation of moments for the case of small intrinsic noise together with extrinsic noise originating from rate constants with a static (time-independent) normally distribution. Roberts \textit{et al.}~\cite{RobertsAssaf_PRE2015} develop a different type of approximation based on the WKB (Wentzel-Kramers-Brillouin) method and the assumption of extrinsic noise originating from rate constants with a static negative binomial distribution. The advantage of the linear-noise approximation methods is the ease with which they can be calculated for systems with a large number of interacting components since the method amounts to solving a Lyapunov equation that can be computed very efficiently~\cite{SchnoerrSanguinettiGrima_JPA2017} while the major disadvantage is that the fluctuating rate parameters can become negative due to the assumption of a normal distribution. In contrast, the WKB method is difficult to extend to more than one variable however the fluctuating rates are positive.

 In this work, we present a novel method based on the linear-noise approximation that is applicable to systems with intrinsic noise together with extrinsic noise originating from rate constants with a static lognormal distribution. It is assumed that the timescale of the extrinsic noise is much longer than that of intrinsic noise, a biologically realistic scenario ~\cite{Raser05} (correlation times for extrinsic fluctuations in \textit{Escherichia coli} is of the order of 40 minutes which corresponds to the cell cycle period~\cite{Rosenfeld05, Hensel12}, while intrinsic processes typically happen on the order of a minute or shorter timescales). Our method is computationally efficient (due to the use of the linear-noise approximation) while maintaining physical realism by enforcing positive fluctuating rate constants. It hence overcomes the disadvantages of the aforementioned existing frameworks (see previous paragraph). It is also the case that the lognormal distribution appears to be ubiquitous in cell biology~\cite{Furusawa, Bengtsson,Peng} and hence it is the obvious choice to characterise the generally non-Gaussian distribution of positive fluctuating rates.   

The paper is divided as follows. In Section II we develop the theory and derive general expressions for the mean, variances and power spectra of fluctuating molecule numbers in a general biochemical system subjected to lognormal extrinsic noise and intrinsic noise. In Section III, the theory is applied to study how extrinsic noise affects: (1) the second moments of protein numbers in a three-stage model of gene expression and an auto-regulatory genetic feedback loop, (2) the information transfer through a simple biochemical system and (3) the robustness of feed-forward motifs.

\section{\label{sec:theory}Theory}
To correctly model the effects of extrinsic noise, the variables describing extrinsic fluctuations
are introduced at the level of the CME. This renders the theory intractable, besides in simplest cases. To overcome this obstacle we propose an asymptotic expansion method relying on three main steps. In Section~\ref{sec:LNA}, we follow van Kampen's system size expansion~\cite{Kampen1961} which is truncated at first order to obtain the well-known linear-noise approximation (LNA)~\cite{Kampen2007} as a function of the time-dependent extrinsic variables.
Subsequently, we assume timescale separation between the fast intrinsic fluctuations and slowly changing extrinsic variables (Section~\ref{sec:TSS}). This
allows us, as a final step, to employ a small noise expansion of the extrinsic stochastic variables and obtain closed-form expressions for the means, variances and power spectra of the biochemical system components in Section~\ref{sec:SNA}. 

\subsection{\label{sec:LNA}LNA with extrinsic variables}
We consider a chemical network with system volume $\Omega$ and $N$ molecular
species with copy numbers $X_{i}(t)$ and concentrations
$x_{i}(t)=X_{i}(t)/\Omega$, where $i \in \{1, \ldots, N\}$.
The chemical reactions are described by $R$ reaction channels with
rate functions $\tilde f_j(\mathbf x, \mb\eta, \Omega)$ with
$j \in \{1, \ldots, R\}$ and
stoichiometric matrix $\mathbf{S} \in \ZZ^{N \times R}$.
External fluctuations in the
rates, that are not included in the microscopic description,
are described by slowly changing stochastic variables $\eta_{k}(t)$,
$k \in \{1, \ldots, M\}$,
where $M$ equals the number of fluctuating parameters $\bar{c}_{k}$,
\begin{align}
  \label{eq:ext_fluct_1}
  \bar{c}_k(t) &= {c}_k \nu_k(t) = c_k \left(1 + \eta_k(t)\right),
\end{align}
such that $\average{\bar{c}_k(t)} = c_k$.
Lognormal coloured noise $\nu_k(t)$ ensures positive rate constants which avoids
spurious production or degradation.
Furthermore, lognormal rather than normal
distributions have been measured for gene expression
rates~\citep{Rosenfeld05,Shahrezaei08}.
The fluctuations around the mean values $c_k$ are then proportional to
$\eta_{k}(t)$.
As shown in Appendix~\ref{sec:noise_construction}, the lognormal variables
$\nu_{k}(t) = \exp\left(\mu_{k}(t) - \frac{1}{2} \epsilon_{k}^2\right)$
with mean $1$ may by constructed from an
Ornstein-Uhlenbeck process for the normal variables $\mu_{k}(t)$ with
standard deviations $\epsilon_{k}$.

The probability that the network is in a state with copy numbers $\mathbf{X}$
and stochastic variables $\mb\eta$ at time $t$ is given by
$P(\mathbf{X,\mb\eta},t)$, and $\Omega  \tilde f_j(\mathbf x, \mb\eta,
\Omega)\diff t$ is the probability that a reaction of type $j$ occurs in time
$\diff t$. The system's state after the reaction is defined by copy numbers
$X_{i}+S_{ij}$ and the value of $\mb\eta$ at that time.

The chemical master equation describing the microscopic system dynamics is then given by
\begin{equation}
  \label{eq:CME}
  \frac{\diff P(\mathbf{X},\mb\eta,t)}{\diff t}=\Omega
  \displaystyle\sum_{j=1}^{R} \left(
  \displaystyle\prod_{i=1}^{N}{E^{-S_{ij}}_{i}}-1 \right)
  \tilde{f}_{j} (\mathbf{x},\mb\eta,\Omega)P(\mathbf{X},\mb\eta,t),
\end{equation}
where $E_{i}$ is a step operator that is defined by the action $E_i^n
g(X_1,...,X_i,...,X_N)=g(X_1,...,X_{i}+n,...,X_N)$, the product of which takes into account
all system states that can evolve to the state given by $\mathbf{X}$ and
$\mb\eta$. The probability flux away from state $(\mathbf{X}, \mb\eta)$ due to
reaction $j$ is given by $ -\Omega \tilde{f}_{j}
(\mathbf{x},\mb\eta,\Omega)P(\mathbf{X},\mb\eta,t)$.

If the system volume is
sufficiently large, the fluctuations of the concentrations $\mathbf{x}$ 
due to a couple of reactions on short timescales are relatively small.
Thus, the rate functions change more or less continuously on a larger timescale.
We can therefore define a
macroscopic limit $\Omega \to \infty$ with the concentrations, transition rates
and deterministic reaction rate equations are given by:
\begin{subequations}
\begin{align}
  \label{eq:macr_limit1}
  \mb\phi &= \lim_{\Omega \to \infty} \mathbf x,
  &\\
  \label{eq:macr_limit2}
  \mathbf{f}(\mb\phi,  \mb\eta) &= \lim_{\Omega \to \infty}
  \tilde{\mathbf{f}} ( \mathbf x,  \mb\eta, \Omega),
  &\\
  \label{eq:macr_limit3}
  \ddfrac[\mb\phi]{t}
  &= \mathbf g(\mb\phi, \mb\eta)
   = \mathbf{S} \mathbf{f}(\mb\phi, \mb\eta) .
\end{align}
\end{subequations}
Following the system size expansion by van Kampen~\cite{Kampen2007}
we relate the microscopic and macroscopic vectors via
\begin{align}
  \label{eq:lna_ansatz}
  \mathbf x &= \mb\phi + \Omega^{-\frac{1}{2}} \mb\xi,
\end{align}
where a new variable $\mb\xi$ denotes the microscopic fluctuations around the
macroscopic concentrations $\mb\phi$. 
It obeys the stochastic differential equation~\citep{Gardiner1990}
\begin{align}
  \label{eq:dbxi}
  \diff\mb\xi(t)
  &= \bA(\mb\eta(t)) \mb\xi(t) \diff t + \bB\left(\mb\eta(t\right)) \diff\mathbf W(t)
\end{align}
with the Wiener process increments $\diff \mathbf W(t)$, Jacobian matrix
$\bA(\mb\eta)$ and diffusion matrix $\bB(\mb\eta)$.

\subsection{\label{sec:TSS}Timescale separation between dynamics of intrinsic and extrinsic fluctuations}

\subsubsection{\label{sec:TSS_system}Integrating a stationary solution}

As Eq.~\eqref{eq:LNA} generally cannot be solved analytically, we assume that the supremum of the 
timescales of intrinsic noise as given by the absolute value of the inverse of the eigenvalues of the Jacobian $\bA$ is much less than the timescale of extrinsic noise. This allows us to split the time axis into intervals$[t_n,t_n+\triangle t]$, on which the extrinsic variables $\mb\eta$ are treated as constants $\mb\eta^n\equiv\mb\eta(t_n)$. 
For well defined stationary solutions we require the existence of a unique macroscopic stationary solution $\mb\phi^{s}(\mb\eta^n)$ of Eq.~\eqref{eq:macr_limit3},
\begin{align}
  \label{eq:phis}
  \mathbf g(\mb\phi^{s}(\mb\eta^n), \mb\eta^n) = 0\,,
\end{align}
and that the Jacobian matrix has only eigenvalues with negative
real part (\textit{i.e.} a stable monotonic system).
The stationary Jacobian and diffusion matrices are
\begin{align}
  \bA(\mb\eta^n) &= \pfrac[\mathbf g]{\mb\phi}(\mb\phi^{s}(\mb\eta^n), \mb\eta^n), \label{eq:a} \\
  \bB\left(\mb\eta^n\right)\bB\left(\mb\eta^n\right)^T
  &= \mathbf{S} \diag\left(\mathbf{f}(\phi^{s}(\mb\eta^n), \mb\eta^n)\right) \mathbf{S}^T . \label{eq:bbt}
\end{align}
Aiming at a stationary solution $\mathbf{x}(t)$ that makes it possible to obtain expressions for the mean, variance and power spectrum, we further follow
the well known \emph{linear-noise approximation}~\cite{Kampen2007}: we linearise the rate equations \eqref{eq:macr_limit3} and use Eqs.~\eqref{eq:lna_ansatz} and \eqref{eq:dbxi} to obtain the linear stochastic differential equation for $t\in [t_n,t_n+\triangle t]$. The stochastic differential equation describing the fluctuations in molecule numbers can be derived by using Eq.~\eqref{eq:lna_ansatz} (with $\phi $ replaced by $\phi^s(\eta^n)$) together with Eq. (5) (with $\eta(t)$ replaced by $\eta^n$) to obtain:
\begin{equation}
  \label{eq:LNA}
  \diff \mathbf{x}(t)=\bA(\mb\eta^n)(\mathbf{x}(t)-\mb{\phi}^{s}(\mb\eta^n))\diff t
  +\frac{1}{\sqrt{\Omega}}\bB(\mb\eta^n)\diff \mathbf{W}(t).
\end{equation}
with the stationary solution \citep{Gardiner1990}:
\begin{subequations}
\label{eq:sol_x}
\begin{align}
  \mathbf x^s(t) &= \mb\phi^s(\mb\eta^n)
  + \frac{1}{\sqrt{\Omega}} \mb\xi(\mb\eta^n, t),
  \\
  \mb\xi(\mb\eta^n, t)
  &= \int_{-\infty}^{t}
  e^{\bA(\mb\eta^n) (t - t')}
  \bB\left(\mb\eta^n\right) \diff\mathbf W(t') .
\end{align}
\end{subequations}

\subsubsection{\label{sec:TSS_mean_var}Calculating mean concentrations and variances}

To evaluate averages and variances of the stationary concentrations $\mathbf x^s$ we denote averaging over intrinsic fluctuations $\mb\xi$ by $\average[i]{~}$
and over the extrinsic variables $\mb\eta$ by $\average[e]{~}$.
The mean concentrations then simplify to
\begin{equation}
  \label{eq:tscale_stx}
  \average[e]{\average[i]{\mathbf x^s}}
  = \average[e]{\mb\phi^s(\mb\eta)}
\end{equation}
where $\average[i]{\mb\xi}$ evaluates to zero on each of the time intervals \cite{Kampen2007}.
In the same way, the covariance matrix of $\mathbf x^s$ in the timescale separation
approximation can be written as
\begin{equation}
\begin{split}
  \label{eq:tscale_Vstx}
  \mathbf V\left(\mathbf x^s\right)
  &= \average{\left(\mathbf x^s - \average{\mathbf x^s}\right) \left(\mathbf x^s - \average{\mathbf x^s}\right)^T}\\
  &= \mathbf V\left(\mb\phi^s\right)
  + \frac{1}{\Omega} \mathbf V\left(\mb\xi\right).
\end{split}
\end{equation}
where $\average{~}$ abbreviates $\average[e]{\average[i]{~}}$ and
we define the covariance matrices of $\mb\phi^s$ and $\mb\xi$ as
\begin{align}
  \label{eq:def:variances}
  \mathbf V\left(\mb\phi^s\right) &=
  \average[e]{\mb\phi^s \mb\phi^{s T}}
  - \average[e]{\mb\phi^s} \average[e]{\mb\phi^{s T}},\\
  \label{eq:def:variances_int}
  \mathbf V\left(\mb\xi\right) &= \average[e]{\average[i]{\mb\xi \mb\xi^T}}.
\end{align}
One can obtain $\mathbf V(\mb\xi)$ algebraically.
For the sake of brevity we anticipate the result
using Eq.~\eqref{eq:def:Gi} and \eqref{eq:def_C}
\begin{align}
  \label{eq:rel:xi_Gi}
  \mathbf V\left(\mb\xi\right) =
  \average[e]{\average[i]{\mb\xi \mb\xi^T}}
  ~=~
  \average[e]{\mathbf G_i(t, t)}
  = \average[e]{\bC(\mb\eta, \mb\eta)}
\end{align}
were the Lyapunov matrix $\bC$ is evaluated at equal times
and we calculate the time correlation function $\mathbf G(t_1, t_2)$
as an intermediate step to calculate the power spectrum in what follows.

\subsubsection{\label{sec:TSS_mean_spectrum}Calculating power spectra via Fourier transformation}

The spectrum matrix
of a stationary stochastic process is connected to
the time correlation function by the \emph{Wiener-Khinchin theorem}
via a Fourier transformation if the time correlation is
sufficiently smooth~\citep{Kampen2007,Gardiner1990},
\begin{align}
  \label{eq:def:spectrum_matrix}
  \bP(\omega)
  = \frac{1}{2\pi} \int_{-\infty}^{+\infty} e^{-i\omega\tau} \mathbf G(\tau,0) \dtau .
\end{align}
We first introduce the time correlation function of a stationary solution
$\mathbf x^s$ (Eq.~\ref{eq:sol_x}) as
\begin{align}
\label{eq:Gt1t2}
  \mathbf G(t_1, t_2) &= \average{
    \bigl(\mathbf x^s(t_1) - \average{\mathbf x^s} \bigr)
    \bigl(\mathbf x^s(t_2) - \average{\mathbf x^s} \bigr)^T
  }
  \\\label{eq:G_as_sum}
  &= \mathbf G_e(t_1, t_2)
  + \frac{1}{\Omega} \average[e]{\mathbf G_i(t_1, t_2)}.
\end{align}
The second equality holds under timescale separation conditions and represents
the sum of the time correlation function of the macroscopic stationary state
\begin{align}
  \mathbf G_e(t_1, t_2) &=
  \average[e]{\mb\phi^s(\mb\eta^1) \mb\phi^{s}(\mb\eta^2)^{T}}
  - \average[e]{\mb\phi^s} \average[e]{\mb\phi^{s T}}
\end{align}
and the time correlation function of the intrinsic noise
subject to slow extrinsic fluctuations
\begin{align}
  \label{eq:def:Gi}
  \mathbf G_i(t_1, t_2) &= \average[i]{
    \mb\xi\left(\mb\eta^1\right) \mb\xi\left(\mb\eta^2\right)^T
  } .
\end{align}
To evaluate Eq.~\eqref{eq:def:Gi} we follow the calculation of the variance
in stationary solution by Gardiner~\citep{Gardiner1990} (having defined $\bA$
as the Jacobian it has opposite sign to the matrix in the reference)
and generalise it by evaluating
the results at different times to obtain
\begin{equation}
  \label{eq:G_and_C}
  \mathbf G_i(t_1, t_2) = e^{\bA(\mb\eta^1) (t_1 - \min(t_1, t_2))} \,\,\,
  \bC\left(\mb\eta^1, \mb\eta^2\right)
  e^{{\bA(\mb\eta^2)}^T (t_2 - \min(t_1, t_2))}
 \end{equation}
where the $\bC$ matrix is defined by the Lyapunov equation
\begin{align}
  \label{eq:def_C}
  \nonumber
  \bA(\mb\eta^1)\,\bC\left(\mb\eta^1, \mb\eta^2\right) + \bC&\left(\mb\eta^1, \mb\eta^2\right) {\bA(\mb\eta^2)}^T \\
  &= - \bB\left(\mb\eta^1\right) {\bB\left(\mb\eta^2\right)}^T .
\end{align}
Exploiting timescale separation, we split the spectrum matrix Eq.~\eqref{eq:def:spectrum_matrix}
into two terms according to Eq.~\eqref{eq:G_as_sum},
\begin{align}
  \label{eq:P_as_sum}
    \bP(\omega)
    = \bP_e(\omega) +
    \frac{1}{\Omega} \average[e]{\bP_{i}(\omega)} .
\end{align}
With Eq.~\eqref{eq:G_and_C} and $\mb\eta^n\equiv\mb\eta(t_n)$ with
$t_{1} = \tau$ and $t_{2} = 0$ we can express $\average[e]{\bP_{i}(\omega)}$
in terms of the matrices
\begin{subequations}
\begin{align}
  \label{eq:simpl_Pi}
  \bR(\omega) &= \average[e]{
    \int_{0}^{\infty}
    e^{-\left(-\bA(\mb\eta^1) + i\omega\right)\tau}
    \bC(\mb\eta^1, \mb\eta^2) \dtau
  },
  \\
  \bR(\omega)^{*T} &= \average[e]{
    \int_{0}^{\infty} \bC(\mb\eta^2,\mb\eta^1)
    e^{\left(\bA(\mb\eta(\tau))^T + i\omega\right)\tau} \dtau
  } .
\end{align}
\end{subequations}
Assuming stationarity of $\mb\eta(t)$ it can be shown that
\begin{align}
  \bR(\omega) + \bR(\omega)^{*T} =
  \int_{-\infty}^{+\infty} e^{-i\omega\tau} \mathbf G_i(\tau,0) \dtau .
\end{align}
Further, we split the Taylor expansion of the Jacobian in two,
\begin{align}
  \bA(\mb\eta) &= \bA^0 + \left(\bA(\mb\eta) - \bA^0\right) ,
  &
  \bA^0 &\equiv \bA(\vect{0}) ,
\end{align}
and summarise the quantities of interest for Eq.~\eqref{eq:P_as_sum}:
\begin{align}
  \label{eq:def_Pe}
  \bP_e(\omega) &= \frac{1}{2\pi} \int_{-\infty}^{+\infty}
  e^{-i\omega\tau}
  \Bigl(\average[e]{\mb\phi^s(\mb\eta^1) \mb\phi^{s}(\mb\eta^2)^{T}}
  \Bigr. \\\nonumber \Bigr.
  & \hspace{3cm}
  - \average[e]{\mb\phi^s} \average[e]{\mb\phi^{s T}}\Bigr)
  \dtau ,
  \\
  \label{eq:def_R}
  \bR(\omega)
  &= \hspace{\templength}\,
  \int_{0}^{\infty} e^{-\left(-\bA^0 + i\omega\right)\tau}
  \times \\\nonumber
  & \hspace{1.4cm}
  \average[e]{e^{\left(\bA(\mb\eta^1) - \bA^0\right)\tau}
  \bC(\mb\eta^1,\mb\eta^2)} \dtau ,
  \\
  \label{eq:def_Pi}
  \average[e]{\bP_{i}(\omega)}
  &= \frac{1}{2\pi} \left( \bR(\omega) + \bR(\omega)^{*T} \right) .
\end{align}

\subsection{\label{sec:SNA}Small noise expansion}
In the third and last step, we expand equations~\eqref{eq:tscale_stx},
\eqref{eq:def:variances}, \eqref{eq:rel:xi_Gi}, \eqref{eq:def_Pe} and
\eqref{eq:def_R} in Taylor series in the $M$ noise variables $\mb\eta$.
To calculate expected values, also of the time dependent integrands
in \eqref{eq:def_Pe} and \eqref{eq:def_R}, we consequently need the
$n$-point time correlation functions of the extrinsic noise variables.
To this end, we need the following results derived in Appendix~\ref{sec:lognormal_corr}.

For smooth functions $y_{k}(\mu_{k})$ of normal stochastic variables
$\mu_{k}$ with $m^{th}$ derivatives $y_{k}^{(m)}(\mu_{k})$, the derived $n$-point time
correlation function reads:
\begin{align}
  \label{eq:general_malakhov}
  \average{y_{1} \dots y_{n}} &=
  \hspace{-.7em}\sum_{\hspace{.7em}|\bd^n| = 0}^{\infty}
  \Biggl(
    \prod_{k=1}^{n} \average{y_{k}^{(m_k)}(\mu_{k})}
    \prod_{
      \substack{ i, j = 1 \\ i < j }
    }^{n}
    \frac{\Delta_{ij}^{d_{ij}}}{d_{ij}!}
  \Biggr)
\end{align}
which generalises a previous result for $n = 2$~\cite{Malakhov1978}.
Each value of $|\bd^n| = \sum d_{ij}$ involves
summation over all index tuples
$\bd^n = (d_{12}, d_{13}, d_{23}, \dots, d_{(n-1)\,n})$.
The 2-point time correlation functions of the normal stochastic variables
are denoted by $\Delta_{ij} = \average{\mu_i \mu_j}$.
We point out the simplicity of this result for lognormal variables
with mean $\average{\nu_{k}} = 1$,
\begin{align}
  \label{eq:timecor_nu}
  &\average{\nu_1 \ldots \nu_n}
  = \exp\Biggl(\sum_{1 \le i < j \le n} \Delta_{ij}\Biggr) .
\end{align}
In the small noise expansion we use the shifted
$\eta_{k} = \nu_{k} - 1$ with the time correlation functions
(see Appendix~\ref{sec:lognormal_corr})
\begin{align}
  \label{eq:timecor_eta}
  \average{\eta_1 \ldots \eta_n} &=
  \sum_{u = \lfloor\frac{n + 1}{2}\rfloor}^{\infty}
  \hspace{-.7em}\sum_{\hspace{.7em}|\bd^n| = u}\raisebox{3pt}{\hspace{-.9em}$'$}\hspace{0.5em}
  ~\prod_{
    \substack{ i, j = 1 \\ i < j }
  }^{n}
  \frac{\Delta_{ij}^{d_{ij}}}{d_{ij}!}
\end{align}
where the prime denotes the restriction of the sum to terms where
for each $j \in \{1, \dots, n\}$ there is a $i < j$ or $k > j$ with
$d_{ij} \neq 0$ or $d_{jk} \neq 0$.
The floor function
$\lfloor \frac{n + 1}{2} \rfloor$ gives the smallest $u$ for which this can be
satisfied.

Selected terms of the final results that we present in the following are
exemplarily evaluated in Appendix~\ref{sec:theory_example} to further clarify
the complex notation.

\subsubsection{Mean}
We define the Taylor expansion of the mean concentrations in
Eq.~\eqref{eq:tscale_stx} with the unusual but beneficial notation
\begin{align}
  \label{eq:small_noise_mean}
  \average{\mathbf x^s(\mb\eta)} = \average{\mb\phi^s(\mb\eta)}
  = \sum_{n=0}^{\infty}
  \sum_{\# \br = n} \mb\phi^s_{(\br)} \average{\eta_{(\br)}}
\end{align}
where the multi-indices
\begin{subequations}
\label{eq:def:multi_r}
\begin{align}
  \br &= \left(r_1, \ldots, r_{n}\right),
  \quad
  \# \br = n,
  \quad
  r_i \in \{1, \ldots, M\}
  \intertext{%
    are sorted by their length $\#\br$ and we denote the Taylor coefficients and
    products of the noise variables as
  }
  \label{eq:def:multi_r_b}
  \mb\phi^s_{(\br)} &\equiv
  \frac{1}{n!} \pfrac{\eta_{r_1}} \ldots \pfrac{\eta_{r_{n}}}
  \left.\mb\phi^s(\mb\eta)\right|_{\mb\eta = \vect{0}} ,
  \\
  \eta_{(\br)} &\equiv \eta_{r_1} \ldots \eta_{r_{n}}.
\end{align}
\end{subequations}
The sum over $\# \br = n$ involves all $\br$ tuples of length $n$.
By inserting Eq.~\eqref{eq:timecor_eta} into Eq.~\eqref{eq:small_noise_mean}
we obtain the final result for the mean concentrations in terms of the
covariances of the independent normal stochastic variables
$\mu_{i}$ with standard deviations $\epsilon_{i}$, that is, their 2-point correlation
functions evaluated at equal times
\begin{align}
  \label{eq:delta1}
  \Gamma_{i j} &\equiv \Delta_{i j}(0) = \delta_{i j} \epsilon_{i}^2
\end{align}
with the Kronecker delta
(see details in Appendix~\ref{sec:small_noise_xs}):
\begin{align}
  \label{eq:result_mean}
  \hspace{-0.25em}
  \average{\mathbf x^s} =
  \sum_{u = 0}^{\infty}
  \sum_{n=0}^{2 u} \sum_{\# \br = n}
  \mb\phi^s_{(\br)}
  \hspace{-.7em}\sum_{\hspace{.7em}|\bd^n| = u}\raisebox{3pt}{\hspace{-.9em}$'$}\hspace{0.5em}
  \,\prod_{
    \substack{ i, j = 1 \\ i < j }
  }^{n}
  \frac{1}{d_{ij}!}
  \Bigl(\Gamma_{r_i r_j}\Bigr)^{d_{ij}}
  \hspace{-0.25em} .
\end{align}
The prime at the sum denotes the same restriction as in Eq.~(\ref{eq:timecor_eta}): the sum is running over all tuples $\bd^n = (d_{12}, d_{13}, d_{23}, \dots, d_{(n-1)\,n})$
with $|\bd^n| = \sum d_{ij} = u$ that obey the condition
that for each $j \in \{1, \dots, n\}$ there is a $i < j$ or $k > j$ such that
$d_{ij} \neq 0$ or $d_{jk} \neq 0$.

\subsubsection{Variance}
With Eq.~\eqref{eq:rel:xi_Gi} we write the matrix
$V(\mb\xi) = \average{\bC(\mb\eta, \mb\eta)}$ in analogy
to $\average{\mathbf x^s} = \average{\mb\phi^s}$.
Because $\bC$ is evaluated at equal times here,
we may expand it according to \eqref{eq:def:multi_r_b}
with Taylor coefficients $\bC_{(\br)}$.
From Eq.~\eqref{eq:result_mean} we then obtain
\begin{align}
  \label{eq:result_Vxi}
  \hspace{-0.25em}
  \mathbf V(\mb\xi) =
  \sum_{u = 0}^{\infty}
  \sum_{n=0}^{2 u} \sum_{\# \br = n}
  \hspace{-0.3em}
  \bC_{(\br)}
  \hspace{-0.3em}
  \hspace{-.7em}\sum_{\hspace{.7em}|\bd^n| = u}\raisebox{3pt}{\hspace{-.9em}$'$}\hspace{0.5em}
  \,\prod_{
    \substack{ i, j = 1 \\ i < j }
  }^{n}
  \frac{1}{d_{ij}!}
  \Bigl(\Gamma_{r_i r_j}\Bigr)^{d_{ij}}
  \hspace{-0.25em} .
\end{align}

The derivation of the series for $\mathbf V(\mb\phi^s)$ is conducted
similarly.  For the expansion of $\mb\phi^s \mb\phi^{s\;T}$ in
Eq.~\eqref{eq:def:variances}, an
ordinary multi-index $\bq = (q_1, q_2)$, $|\bq| = q_1 + q_2$
with integer $q_1, q_2 \ge 1$ is used to define the lengths of
two multi-indices $\br^1$ and $\br^2$ (see Eq.~\ref{eq:def:multi_r})
and finally
\begin{equation}
  \label{eq:result_Vphis}
  \mathbf V(\mb\phi^s) =
  \sum\limits_{u = 1}^{\infty}
  \sum\limits_{n = 2}^{2 u}
  \sum\limits_{|\bq| = n}
  \hspace{-1.2em}\sum\limits_{\hspace{1.2em}\# \br^1 = q_1}
  \hspace{-1.2em}\sum\limits_{\hspace{1.2em}\# \br^2 = q_2}
  \mb\phi^s_{(\br^1)} \mb\phi^{s\,T}_{(\br^2)}
  \hspace{-.7em}\sum_{\hspace{.7em}|\bd^n| = u}\raisebox{3pt}{\hspace{-.9em}$''$}\hspace{0.5em}
  \,\prod_{
    \substack{ i, j = 1 \\ i < j }
  }^{n}
  \frac{1}{d_{ij}!}
  \Bigl(\Gamma_{f^2_i f^2_j}\Bigr)^{d_{ij}} .
\end{equation}
The doubly primed sum (see Appendix~\ref{sec:small_noise_var}) denotes the
restriction of the summation by one additional
condition: there is at least one $d_{ij} \neq 0$ for which $i \le q_1$ and
$j > q_1$ since all other terms cancel in the
subtraction in Eq.~\eqref{eq:def:variances}.
For later use, we define
the index functions $f^{k}_{i}(\br^1, \dots, \br^k)$
for a generalised set of multi-indices $\br^1, \dots, \br^k$
with $\# \br^i = q_i$ and $\bq = (q_1, \dots, q_k)$,
\begin{align}
  \label{eq:delta2}
  f^{k}_{i} &= \begin{cases}
    r_i^1 &\text{if } \,1 \le i \le q_1
    \\
    r_{(i-q_1)}^2 &\text{if } q_1 < i \le q_1 + q_2
    \\
    \;\vdots &\;\vdots
    \\
    r_{(i-|\bq|+q_k)}^k &\text{if } q_1 + \dots + q_{k-1} < i \le |\bq|
  \end{cases}
\end{align}
that we have used in Eq.~\eqref{eq:result_Vphis} to refer to $\br^1$ and $\br^2$.

\subsubsection{Power spectrum}
In Eq.~\eqref{eq:def_Pe} for the spectrum matrix $\bP_e(\omega)$ the
integral from $-\infty$ to $0$ is the complex conjugate of the integral from
$0$ to $\infty$.  Thus the small noise expansion proceeds in complete
analogy to $\mathbf V\left(\mb\phi^s\right)$ in Eq.~\eqref{eq:def:variances}
with the result in Eq.~\eqref{eq:result_Vphis} when the product of
covariances $\Gamma_{f^2_i f^2_j}$ is
replaced by the Fourier transform plus its complex conjugate
\begin{align}
  \label{eq:integrals_Pe}
  \frac{1}{2\pi} \int_{0}^{\infty} e^{- \ii \omega \tau}
  \prod_{
    \substack{ i, j = 1 \\ i < j }
  }^{n}
  \frac{1}{d_{ij}!}
  \Bigl(\Delta_{f^2_i f^2_j}(t_i - t_j)\Bigr)^{d_{ij}}
  \dtau
\end{align}
where the 2-point time correlation functions are
\begin{align}
  \label{eq:delta}
  \Delta_{ij}(t_i - t_j) \equiv \average{\mu_{i}(t_i) \mu_{j}(t_j)}
  = \Gamma_{i j}
  e^{- K_{i} | t_i - t_j |}
\end{align}
with inverse correlation times $K_{i}$.
We finally obtain
\begin{multline}
  \label{eq:result_Pe}
  \bP_e(\omega)
  =
  \sum\limits_{u = 1}^{\infty}
  \sum\limits_{n = 2}^{2 u}
  \sum\limits_{|\bq| = n}
  \hspace{-1.2em}\sum\limits_{\hspace{1.2em}\# \br^1 = q_1}
  \hspace{-1.2em}\sum\limits_{\hspace{1.2em}\# \br^2 = q_2}
  \mb\phi^{s}_{(\br^1)} {\mb\phi^{s}_{(\br^2)}}^{\,T}\;
  \times\\
  \hspace{-.7em}\sum_{\hspace{.7em}|\bd^n| = u}\raisebox{3pt}{\hspace{-.9em}$''$}\hspace{0.5em}
  \frac{
    \pi^{-1} \, \Theta(\bd^n, \br^1)
  }{
    \omega^2 + \Theta^2(\bd^n, \br^1)
  }
  \prod_{
    \substack{ i, j = 1 \\ i < j }
  }^{n}
  \frac{1}{d_{ij}!}
  \Bigl(\Gamma_{f^2_i f^2_j}\Bigr)^{d_{ij}}
\end{multline}
with $\Theta(\bd^n, \br^1)$ derived from integral \eqref{eq:integrals_Pe}
in Appendix~\ref{sec:small_noise_Pe},
\begin{align}
  \label{eq:integrals_Pe_final}
  \Theta(\bd^n, \br^1)
  &= \sum_{i=1}^{\# \br^1}
  \hspace{-2em}\sum_{\hspace{2em}j = \# \br^1 + 1}^{n}
  d_{ij} \, K_{r^1_i} \,.
\end{align}

To calculate the second spectrum matrix under influence of
extrinsic fluctuations, $\average{\bP_{i}(\omega)}$ in Eq.~\eqref{eq:def_Pi},
the $\bR(\omega)$ matrix in Eq.~\eqref{eq:def_R} requires
expanding.  First, we expand the Jacobian matrix $\bA(\mb\eta^1)$ at time $t_1$
in the same way as $\mb\phi^s(\mb\eta)$ in Eq.~\eqref{eq:small_noise_mean}
with Taylor coefficients $\bA_{(\br)}$
using the multi-index $\br$ from Eq.~\eqref{eq:def:multi_r}.
The Lyapunov matrix $\bC(\mb\eta^1, \mb\eta^2)$ (Eq.~\ref{eq:def_C}) is
evaluated with respect to two arguments corresponding to different times.
Thus, the Taylor expansion requires a second multi-index
\begin{align}
  \label{eq:def:multi_s}
  \bs &= \left(\sigma_1, \ldots, \sigma_a\right) ,\!
  &
  \# \bs &= a ,\!
  &
  \sigma_i &\in \{1, 2\} ,
\end{align}
such that index $\sigma_{i}$ specifies the argument with respect to which the
derivative is taken for the
Taylor coefficients and to which the components $\eta_{i}$ belong,
\begin{align}
  \label{eq:taylor_C_s}
  \bC_{(\br,\bs)} &\equiv
    \frac{1}{a!}
    \pfrac{\eta_{r_1}^{\sigma_1}}
    \ldots
    \pfrac{\eta_{r_a}^{\sigma_a}}
    \left.\bC(\mb\eta^1, \mb\eta^2)\right|_{\mb\eta^1 = \mb\eta^2 = \vect{0}} ,
  \\
  \eta_{(\br,\bs)} &\equiv \eta_{r_1}^{\sigma_1}
    \ldots \eta_{r_a}^{\sigma_a} .
\end{align}
The Taylor expansion of $\exp\left(\left(\bA(\mb\eta^1) - \bA^0)\right)\tau\right)$
in Eq.~\eqref{eq:def_R} in $c$'th order involves $c$ different Taylor
coefficients of $\bA$.  To distinguish them, we use sets of
multi-indices $\br^1, \dots, \br^c$ with an ordinary multi-index
\begin{align}
  \label{eq:def:multi_q}
  \bq &= (q_1, \dots, q_c) ,
  &
  |\bq| &= q_1 + \dots + q_c
\end{align}
and $\# \br^i = q_i \ge 1$ for all $i = (1, \dots, c)$.
The result of the expansion and integration of $\bR(\omega)$ in
Appendix~\ref{sec:small_noise_Pi} reads
\begin{align}
  \nonumber\hspace{-0.6em}
  \bR(\omega)
    &=
    \sum\limits_{u=0}^{\infty}
    \sum\limits_{n=0}^{2u}
    \sum\limits_{a=0}^{n}
    \hspace{-1.4em}\sum\limits_{\hspace{1.4em}|\bq| = n - a}
    \hspace{-1.2em}\sum\limits_{\hspace{1.2em}\# \br^1 = q_1}
    \hspace{-0.5em}\ldots
    \hspace{-1.6em}\sum\limits_{\hspace{1.6em}\# \br^c = q_c}
    \hspace{-1.9em}\sum\limits_{\hspace{1.9em}\# \br^{c+1} = a}
    \hspace{-0.5em}\sum\limits_{\hspace{0.5em}\# \bs = a}
  \\&\hspace{-1.4em}\times
  \nonumber
    \hspace{-.7em}\sum_{\hspace{.7em}|\bd^n| = u}\raisebox{3pt}{\hspace{-.9em}$'$}\hspace{0.5em}
    \frac{
      1
    }{
      \Bigl(- \bA\!^0 + \theta(\bd^n, |\bq|, \br^{c+1}, \bs) + i \omega\Bigr)^{c + 1}
    }
  \\&\hspace{-1.4em}\times
    \bA_{(\br^1)} \ldots \bA_{(\br^c)} \bC_{(\br^{c+1},\bs)}
    \prod_{
      \substack{ i, j = 1 \\ i < j }
    }^{n}
    \frac{1}{d_{ij}!}
    \Bigl(\Gamma_{f^{c+1}_i f^{c+1}_j}\Bigr)^{d_{ij}}\hspace{-1ex}
  \label{eq:result_R}
\end{align}
where the sum $\sum_{|\bq| = n - a}$ is carried out over all possible
$c$, $(q_1, \dots, q_c)$, $q_i \ge 1$ with $q_1 + \dots + q_c + a = n$ and
\begin{align}
  \label{eq:integrals_Pi_final}
  \hspace{-0.7em}
  \theta(\bd^n, |\bq|, \br^{c+1}, \bs)
  = \!\!\!\sum_{
    \substack{ i, j = 1 \\ i < j }
  }^{n}
  d_{ij} \, K_{r^{c+1}_{j-|\bq|}} \, \beta_{ij}(|\bq|) ,
  \hspace{-0.3em}
  \\
  \label{eq:Pi_beta_gamma}
  \beta_{ij}(x) = \begin{cases}
    1 &\text{if } i \le x < j \text{ and } \sigma_{(j - x)} \neq 1,
    \\
    1 &\text{if } x < i < j \text{ and } \sigma_{(j - x)} \neq \sigma_{(i - x)},
    \\
    0 &\text{else} .
  \end{cases}
\end{align}

Finally, we add the complex conjugate to $\bR(\omega)$ and divide by $2\pi$
to obtain the spectrum matrix $\average{\bP_{i}(\omega)}$ as described by
Eq.~\eqref{eq:def_Pi}.

\subsubsection{\label{sec:automated}Automated sum evaluation}

While the derivation of the presented results involves a double expansion and a
rather complicated notation, its strength lies in the closed-form expressions
for the mean, variance and power spectrum of a biochemical system under time
scale separation conditions.  Stochastic modelling of the underlying chemical
master equation with extrinsic fluctuations requires an extension of the
Gillespie algorithm~\cite{Gillespie2007,Shahrezaei08} and it is usually
difficult or inefficient to achieve accuracy at different timescales.
A fast automated evaluation of the closed-form expressions allows for a
systematic approach to analyse the effect of extrinsic fluctuations as a
function of different parameters.

The automated sum evaluation has been implemented with the SymPy library for
symbolic mathematics~\cite{SymPy} in the \verb!ext_noise_expansion! program that
can be obtained from \url{github.com}~\cite{Bastian2013}.
Simple systems can be partially or fully
evaluated symbolically while larger systems may require numerical parameter values.
The limiting step is the calculation of the stationary state $\mb\phi^{s}$ in
Eq.~\eqref{eq:phis} as an analytical function of the extrinsic variables.
For this task, an external solver adapted to the system of interest may be used.
All sums are formally evaluated before inserting the Taylor coefficients and
subsequent term simplification.
Taylor coefficients for $\mb\phi^{s}$, $\bA$ and $\bB$ are directly calculated
using memoisation to avoid multiple evaluations.
Taylor coefficients of $\bC\left(\mb\eta^1, \mb\eta^2\right)$ from the Lyapunov
equation Eq.~\eqref{eq:def_C} are obtained by expansion and recursive
coefficient comparison (this leads to Lyapunov equations for each coefficient
that can be constructed explicitly for any order~\cite{Bastian}).
More details are presented in Appendix~\ref{sec:theory_example}.

\subsection{Stochastic simulations}
Stochastic simulations were performed using the Extrande algorithm by Voliotis \textit{et al.}~\cite{Voliotis15} implemented in \verb!C++11!. Each stochastic simulation data point in Figures~\ref{fig:linearmodel} and~\ref{fig:negfeedback} was obtained from 100 trajectories of $10^8$ seconds each.

\subsection{\label{sec:Optimisation}Optimisation routine}
All optimisations were performed using the L-BFGS-B algorithm (\verb!fmin_l_bfgs_b()!) provided in the SciPy library~\cite{SciPy}. Parameters were generated from a uniform distribution with bounds $[0.0001,20]$.


\section{\label{sec:results}Results}
Here, we will present several applications of the theory to study features of noisy biochemical networks. We first introduce a linear model of gene expression to establish the (limits of) validity of our method and quantify the effects of extrinsic noise in different system parameters on intrinsic and extrinsic cell-to-cell variability. We go on to study the potential of negative feedback control to suppress gene expression noise in the presence of extrinsic noise. In a third example, we apply the notion of mutual information to a simple biochemical system to analyse how extrinsic fluctuations affect a network's ability to relay information. Finally, we present an efficient multi-objective optimisation scheme that combines our analytical framework with deterministic dynamics to obtain optimal network topologies and parameters for feed-forward loops.

\subsection{\label{sec:linear}Nonlinear effects of extrinsic fluctuations on a linear model of gene expression}
\begin{figure*}
  \includegraphics[width=\textwidth]{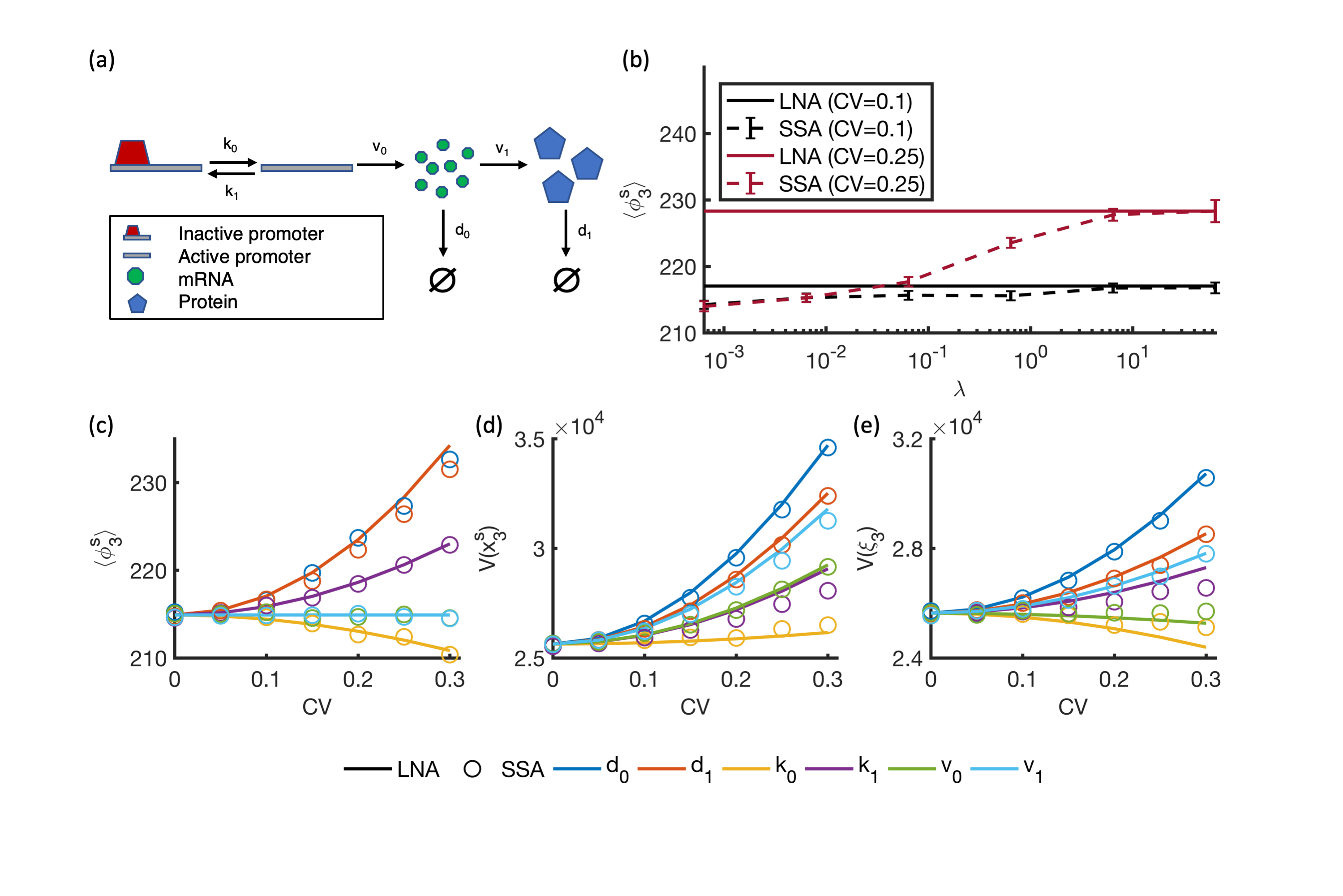}
  \vspace{-2cm}
  \caption{\footnotesize \label{fig:linearmodel} (a) Model of gene expression where the promoter can switch between the active and inactive state. (b) Validity of the timescale separation condition depends on the ratio of extrinsic and intrinsic timescales $\lambda$ and the magnitude of the extrinsic noise denoted by the CV (here in parameter $d_0$). The mean number of protein $A$ at steady state, denoted by $\average{\phi_3^s}$, with black lines corresponding to extrinsic noise with $CV=0.1$, and red lines corresponding to extrinsic noise with $CV = 0.25$. Solid lines are analytical predictions (denoted by ``LNA"), dashed lines are stochastic simulation results (denoted by ``SSA") from the Extrande algorithm~\cite{Voliotis15}. 95$\%$ confidence intervals for the mean were calculated following Bran{\v{c}}{\'\i}k \textit{et al.} in~\cite{Branvcik16}. (c) Effect of extrinsic noise in different parameters on the mean number of protein $A$. Analytical predictions for fluctuations in parameter: $d_0$ (blue line, behind the red line), $d_1$ (red line), $k_0$ (yellow line), $k_1$ (purple line), $v_0$ (green line, behind the light blue line), and $v_1$ (light blue line). We calculate terms up to sixth order in $\epsilon$ (substitute $u = 3$ in Eq.~\eqref{eq:result_mean}). Circles denote stochastic simulation results. (d) Effect of extrinsic noise in different parameters on the total variance of protein $A$, given by $V(x^s_3)$ (see Eq. \ref{eq:tscale_Vstx}). We calculate terms up to second order in $\epsilon$ (substitute $u = 1$ in Eqs.~\eqref{eq:result_Vphis}, \eqref{eq:result_Vxi}). Line and circle colours correspond to extrinsic noise in the same parameters as in (c). (e) Effect of extrinsic noise in different parameters on the intrinsic variance of protein $A$, given by $V(\xi_3)$ (see Eq. \ref{eq:tscale_Vstx}). We calculate terms up to second order in $\epsilon$ (substitute $u = 1$ in Eq.~\eqref{eq:result_Vxi}). We use the dual reporter technique ~\cite{Elowitz02} to compute estimates of the intrinsic variance from stochastic simulations. Line and circle colours correspond to extrinsic noise in the same parameters as in (c).}
\end{figure*}

We start by verifying the validity of our method on the well-known three-stage model of gene expression~\cite{Raser04} by introducing coloured noise in various system parameters and studying its effect on the mean and variance of protein numbers. The two-stage gene expression model shown in Figure~\ref{fig:linearmodel}(a) includes the dynamics of active promoter molecules $D$, inactive promoter molecules $D^*$, mRNA molecules $M$, and protein molecules $A$ (note that this notation is unrelated to any terms of the same name in Section~\ref{sec:theory}), and is described by the following reactions
\begin{equation}
\begin{split}
\label{eq:GE}
D^*\underset{k_1}{\stackrel{k_0}{\rightleftharpoons}}D\overset{v_0}\rightarrow& D + M, \,\,\,\,\,\,\,\,\,\,  M\overset{d_0}{\rightarrow}\emptyset, \\
M \overset{v_1}\rightarrow& M+A, \,\,\,\,\,\,\,\,\,\, A\overset{d_1}{\rightarrow}\emptyset.
\end{split}
\end{equation}
We will now show how to write down the LNA as in Eq.~\eqref{eq:LNA} for extrinsic noise in one of the reaction rates of this system. To introduce extrinsic fluctuations $\eta$ into the rate function, we multiply a rate constant, e.g. $d_1$, by a lognormally distributed stochastic variable $\nu = 1+ \eta$

\begin{equation}
\label{eq:d1extnoise}
\bar{d_1}(t)=d_1 \nu (t) = d_1 (1 + \eta(t)).
\end{equation}
The lognormally distributed variable can be constructed from a normal stochastic variable $\mu$ with mean zero, variance $\epsilon^2$, and correlation time $\tau = K^{-1}$ defined by the Ornstein-Uhlenbeck process

\begin{equation}
\label{eq:dmu}
\diff \mu(t) = -K \mu(t) + \sqrt{2 K} \epsilon \diff W(t).
\end{equation}
The shifted lognormal stochastic variable $\eta$ is then given by

\begin{equation}
\eta(t)=\exp \Big( \mu(t) - \frac{1}{2} \epsilon^2 \Big) - 1,
\end{equation}
and we define the magnitude of the extrinsic noise as the coefficient of variation

\begin{equation}
\label{eq:CV}
CV = \sqrt{\exp(\epsilon^2)-1},
\end{equation}
such that the average of the lognormal variable $\nu(t) = 1$ and its standard deviation is equal to Eq.~\eqref{eq:CV}. To be able to write down the LNA in the way of Eq.~\eqref{eq:LNA}, we specify the Jacobian matrix $\bA$ and the diffusion matrix $\bB\bB^T$. The deterministic rate equations for the macroscopic concentrations $\mb\phi$ of the molecular species (where $\phi_1 = D$, $\phi_2 = M$, and $\phi_3 = A$) are
\begin{equation}
\label{eq:GEode}
\frac{\diff\mb{\phi}}{\diff t} = \begin{psmallmatrix} -k_1 \phi_1 + k_0(1-\phi_1) \\ v_0 \phi_1 - d_0 \phi_2 \\ v_1 \phi_2 - d_1(1+\eta) \phi_3 \end{psmallmatrix}.
\end{equation}
The dynamics of $D^*$ can be eliminated as the total promoter concentration is conserved $D+D^*=1/\Omega$. One of the conditions for the LNA is that it is valid in the limit of large system size $\Omega$. However, as the current system is linear (i.e. it contains no bimolecular reactions), the LNA will give the exact expressions for the mean and variance independent of the value of $\Omega$~\cite{Grima15}. For this reason, we have chosen $\Omega = 1$ here.

The Jacobian as defined in Eq.~\eqref{eq:a} is then given by
\begin{equation}
\label{eq:GEjac}
\bA = \begin{psmallmatrix} -(k_0+k_1) & 0 & 0 \\ v_0 & -d_0 & 0 \\ 0 & v_1 & -d_1(1+\eta) \end{psmallmatrix},
\end{equation}
and the diffusion matrix as defined in Eq.~\eqref{eq:bbt} for the system is given by
\begin{equation}
\label{eq:GEdiff}
\bB\bB^T = \begin{psmallmatrix} k_1 \phi_1^s + k_0(1 - \phi_1^s) & 0 & 0 \\ 0 & v_0 \phi_1^s + d_0 \phi_2^s & 0 \\ 0 & 0 & v_1 \phi_2^s  + d_1(1+\eta) \phi_3^s \end{psmallmatrix},
\end{equation}
where $\phi_i^s$ are the steady-state concentrations of the active promoter, mRNA, and protein molecules, respectively. The procedure is the same for extrinsic noise in any other system parameter. In summary, for the three-stage model of gene expression the SDE studied in Eq.~\eqref{eq:LNA} is defined by the steady state concentrations $\mb{\phi^s}$, the Jacobian as given in Eq.~\eqref{eq:GEjac}, the diffusion matrix as given in Eq.~\eqref{eq:GEdiff}, and the extrinsic noise $\nu(t)$ in $d_1$ as defined by Eq.~\eqref{eq:d1extnoise} and Eq.~\eqref{eq:dmu}.

The reaction rates used to obtain the results in Figure~\ref{fig:linearmodel} are shown in Table~\ref{tab:Linearparameters} and are representative for gene expression in mammalian cells as determined by Schwanh\"{a}usser \textit{et al.} and Suter \textit{et al.}~\cite{Schwanhausser11, Suter11}. Schwanh\"{a}usser \textit{et al.} experimentally determined transcription and translation rates for over 5\,000 genes in mammalian cells~\cite{Schwanhausser11}. We selected the mode of these parameter distributions as parameter values for our linear model of gene expression. To ensure relatively fast intrinsic timescales, we chose a protein degradation and mRNA degradation rate associated with a gene that was classified as having both unstable protein and mRNA (see Figure 5 in ~\cite{Schwanhausser11}), while also enforcing that the protein lifetime is much longer than the mRNA lifetime. Promoter activation and deactivation rates were chosen as the average over various mammalian genes as measured by Suter \textit{et al.}~\cite{Suter11}. 

In the three-stage gene expression model, the three intrinsic timescales are $1/(k_0+k_1)$, $1/d_0$, and $1/d_1$, corresponding to the lifetimes of the molecular species $D$, $M$, and $A$ respectively. The timescale of the extrinsic noise process is $\tau=K^{-1}$ as defined in Eq.~\eqref{eq:dmu}. The ratio of the extrinsic and intrinsic timescales is then 

\begin{equation}
\lambda = \tau/\max(\tau_{\mathrm{int}}),
\label{eq:lambda}
\end{equation}
where $\max(\tau_{\mathrm{int}})$ is the longest intrinsic timescale in the model, here $1/d_1 = 15\,625$ s. To see for which values of $\lambda$ our method gives good results, we simulate the model in Figure~\ref{fig:linearmodel} with the mRNA degradation rate $d_0$ subject to extrinsic noise with various values of $\tau$ using the Extrande algorithm. Figure~\ref{fig:linearmodel}(b) shows that for longer extrinsic correlation times $\tau$ (corresponding to larger values of $\lambda$), the stochastic simulations (denoted ``SSA'' in Figure~\ref{fig:linearmodel}) approach the analytically calculated mean number of protein $A$, given by $\average{\phi_3^s}$ (see Eq.~\eqref{eq:result_mean}). For both extrinsic noise magnitudes $CV = 0.1$ and $CV = 0.25$, timescale separation conditions are satisfied for $\lambda \approx 10$ and we are able to accurately predict the mean number of proteins. This corresponds to an extrinsic timescale of $\tau = 10^5$ s, which is of the same order of magnitude as the period of the mammalian cell cycle (approximately 27.5 h~\cite{Schwanhausser11}).

\begin{table}
\caption{\label{tab:Linearparameters}Parameter values used for the gene expression model.}
\begin{ruledtabular}
\begin{tabular}{cc}
parameter & value ($s^{-1}$)  \\
\hline
$k_0$ & 0.00085   \\
$k_1$ & 0.0017   \\
$v_0$ & 0.00028 \\
$v_1$ & 0.028 \\
$d_0$ & 0.00019 \\
$d_1$ & 0.000064 \\
\end{tabular}
\end{ruledtabular}
\end{table}

Noise in different parameters can affect the system in different ways. Figure~\ref{fig:linearmodel}(c) shows that the mean protein number can decrease ($k_0$), increase ($d_0$, $d_1$, $k_1$), or remain constant ($v_0$, $v_1$) in response to increasing extrinsic noise magnitude (CV). The total variance in protein $A$, denoted by $V(x^s_3)$ (see Eq.~\eqref{eq:tscale_Vstx}), always increases when a system is affected by extrinsic noise (Figure~\ref{fig:linearmodel}(d)). Extrinsic noise in a parameter that controls the lifetime of molecular components, such as the mRNA or protein degradation rate, might cause intrinsic and extrinsic timescales to mix and thus have non-trivial effects. This can be seen, for example, when extrinsic fluctuations are applied to the promoter activation rate $k_0$, causing a decrease in the intrinsic variance, denoted by $V(\xi_3)$  (Figure~\ref{fig:linearmodel}(e)). The potential of extrinsic fluctuations to decrease the intrinsic noise in gene expression was already noted by Shahrezaei \textit{et al.}~\cite{Shahrezaei08}.

\subsection{\label{sec:feedback}Regulated gene expression}
\begin{figure*}
  \includegraphics[width=\textwidth]{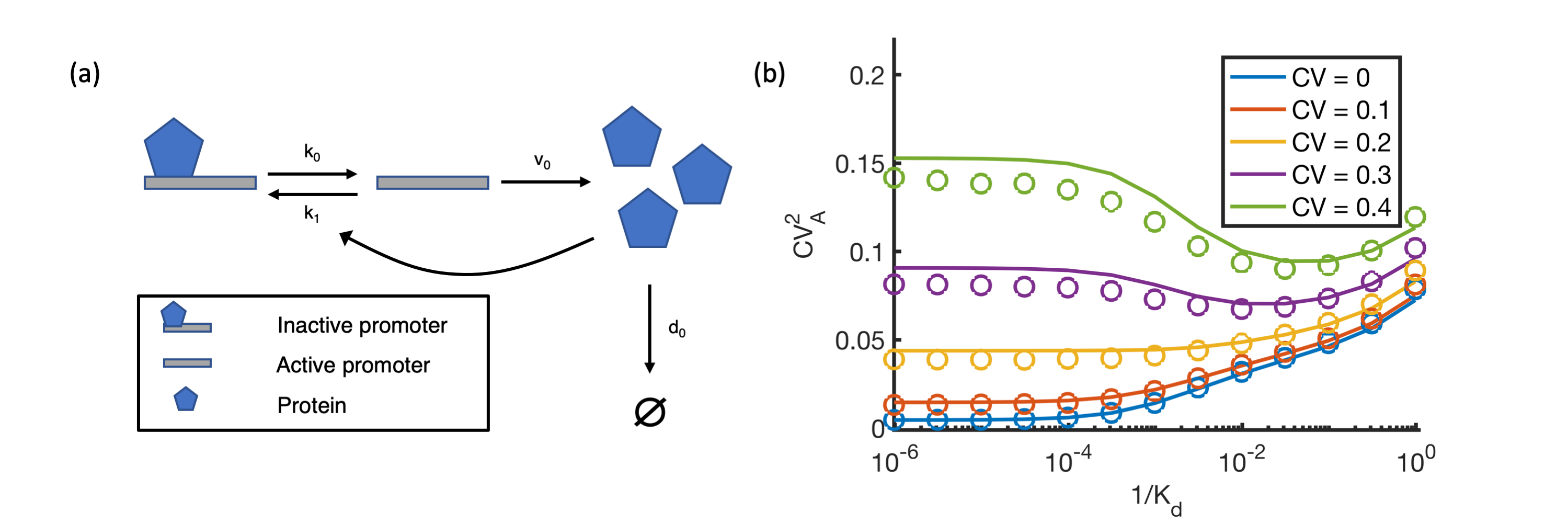}
  \caption{\label{fig:negfeedback} (a) Model of autoregulatory gene expression. The feedback strength $1/K_d = k_1/(k_0 \Omega)$ is determined by the binding affinity of the protein to the promoter. (b) Analytical predictions (solid lines) and stochastic simulation results (circles) for the $CV_{A}^2$ of the number of molecules of protein A as a function of feedback strength $1/K_d$ for extrinsic noise in protein production rate $v_0$ ranging in magnitude from no extrinsic noise ($CV=0$, blue lines and circles) to strong extrinsic noise ($CV=0.4$, green lines and circles). The analytical solution for the mean was calculated up to sixth order in $\epsilon$ (substitute $u = 3$ in Eq.~\eqref{eq:result_mean}), variances were calculated up to second order in $\epsilon$ (substitute $u = 1$ in Eqs.~\eqref{eq:result_Vphis}, \eqref{eq:result_Vxi}).}
\end{figure*}

Feedback control is often proposed as a possible strategy to reduce gene expression noise. Several studies have confirmed that such strategies are indeed capable of reducing the variability in protein concentrations as well as influencing the number of modes of the protein number distribution~\cite{Dublanche06, Zhang09, Lestas10, Toni13, Gronlund13, Oyarzun14,Cao18}. To study the potential of negative feedback to reduce protein noise in a system subject to extrinsic noise, we adapt the genetic feedback loop model proposed by Grima \textit{et al.}~\cite{Grima12}. In this model, the mRNA dynamics are omitted, which is a valid assumption in the absence of translational bursting~\cite{Shahrezaei08analytical, Paulsson05}. The genetic feedback model contains a negative feedback loop where the gene product can bind to the promoter, thus preventing protein production (Figure~\ref{fig:negfeedback}(a)). This system consists of the following reactions
\begin{equation}
\label{eq:FB}
D\overset{v_0}{\rightarrow}D+A\underset{k_0}{\stackrel{k_1}{\rightleftharpoons}} D^*, \,\,\,\,\,\,\,\,\,\,\,\,\,\,\,\,\,\,\,\,\,\,\, A\overset{d_0}{\rightarrow}\emptyset.
\end{equation}
Here, $D$ denotes the unbound promoter,  $D^*$ the bound, inactive promoter, and $A$ the protein. The values for the parameters $k_0$, $v_0$, and $d_0$ (Table~\ref{tab:fbparameters}) were chosen so that they are consistent with those in the linear gene expression model, as follows. Since the mRNA dynamics are omitted, $d_0$ now refers to the protein degradation rate. To ensure the same steady-state species concentrations as in the unregulated gene expression model, we choose the protein production rate as 
\begin{equation}
v_0=\phi_2^s v_1,
\end{equation}
where $\phi_2^s$ refers to the macroscopic steady-state concentration of mRNA in the linear gene expression model, respectively, and $v_1$ is equal to the parameter value of the translation rate of the linear gene expression model as stated in Table~\ref{tab:Linearparameters}. The probability that the binding reaction occurs in a small time interval is proportional to $k_1/\Omega$, where $k_1$ is the protein-DNA binding rate constant and $\Omega$ is the cell volume (approximately 2 picolitres~\cite{Lang92}). We vary the binding rate $k_1$ of protein $A$ to the promoter over a range of biologically relevant specificities~\cite{Bar11} and define the feedback strength as the inverse of the non-dimensional dissociation constant $K_d$

\begin{equation}
\label{eq:fbstrength}
\frac{1}{K_d} = \frac{k_1}{k_0 \Omega}.
\end{equation}
We note that for very small values of $1/K_d$, the system is weakly non-linear since protein binding becomes a rare event compared to promoter activation.
\begin{table}
\caption{\label{tab:fbparameters}Parameter values used for the autoregulatory gene expression model.}
\begin{ruledtabular}
\begin{tabular}{cc}
parameter & value ($s^{-1}$)  \\
\hline
$k_0$ & 0.00085   \\
$k_1/\Omega$ & $8.3 \times 10^{-10} - 8.3\times 10^{-4}$   \\
$v_0$ & 0.014 \\
$d_0$ & 0.000064 \\
\end{tabular}
\end{ruledtabular}
\end{table}

In a negative feedback system with no extrinsic noise, it is well known that the noise increases with $1/K_d$~\cite{Marquez10, Voliotis12, Singh09}. This is  confirmed in Figure~\ref{fig:negfeedback}(b), where the blue lines and circles ($CV=0$) show that the noise in the protein population, quantified by its coefficient of variation squared $CV_A^2$, increases with increasing negative feedback strength. We consider the case of extrinsic noise affecting the protein production rate $v_0$ and show in Figure~\ref{fig:negfeedback}(b) how this changes the noise in the number of proteins $A$. While adding extrinsic noise increases protein noise, as the extrinsic noise magnitude increases beyond a threshold ($CV \sim 0.3$) the magnitude of protein noise does not increase monotonically with $1/K_d$. Rather, it has a minimum at $1/K_d \approx 10^{-1}$. This finding is supported by experimental data that shows that negative autoregulation mechanisms are able to negate the effects of slow extrinsic noise~\cite{Hensel12}. In the absence of extrinsic noise, the variance of protein numbers only has an intrinsic component. The magnitude of intrinsic noise is influenced the average expression level of molecular species and the response time of the system, which is the time it takes for any initial perturbation to decay and the system to return to its equilibrium. In one instance, protein expression levels decrease monotonically with increasing negative feedback strength, and smaller molecular numbers are associated with higher intrinsic noise. In the second, negative feedback is known to speed up the response time of a system, leading to the attenuation of protein noise ~\cite{Rosenfeld02}. These effects can be seen from the blue line and circles in Figure~\ref{fig:negfeedback}(b) indicate that smaller values of $1/K_d$ result in a less noisy system when the system is not subject to extrinsic noise, which implies that the decrease of intrinsic noise due to a faster response time cannot compensate for the increase in noise resulting from smaller protein levels. The extrinsic contribution to the protein noise is also a function of the response time of the system~\cite{Singh09}, which causes the extrinsic component of the protein noise to decrease monotonically with increasing feedback strength. As the $CV$ of the extrinsic noise source increases above $0.3$ (Figure~\ref{fig:negfeedback}(b), purple and green line and circles), the reduction in extrinsic noise is larger than the increase in intrinsic noise up until $1/K_d \approx 10^{-1}$. As the negative feedback strength increases further, the reduction of extrinsic noise is negated by an increase in intrinsic noise due to decreasing protein levels, resulting in an increase of $CV_A^2$.

\subsection{\label{sec:MI}Signal transduction and extrinsic noise}
Cells are embedded in highly fluctuating environments. It is vital for biological systems that they can sense external stimuli and process this information in order to adapt to their environment accordingly. Information theoretic approaches have, for example, been applied to biological systems to address the question of how well a network subject to biochemical noise is able to transmit information that arrives at cell receptors into the intracellular environment.

\begin{figure*}
  \includegraphics[width=\textwidth]{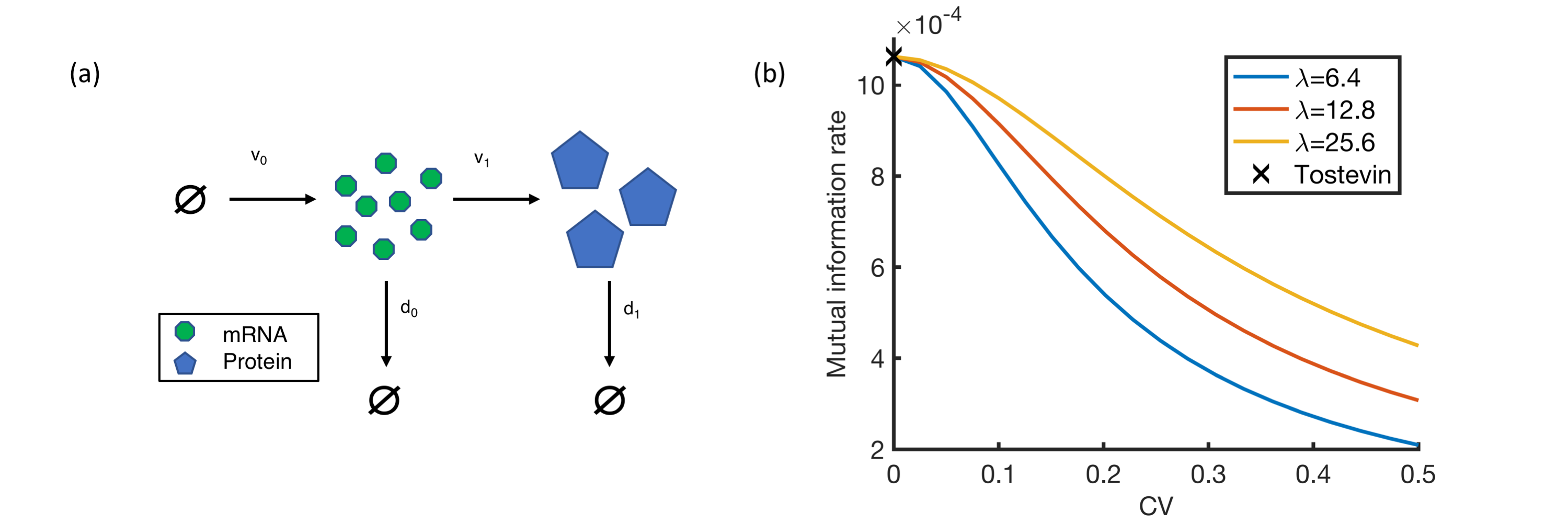}
  \caption{\label{fig:MI}(a) Model of the two-stage model of gene expression, where information about the concentration of mRNA (input) is transduced to the concentration of protein (output). (b) Mutual information rate (solid lines) between input and output variables as a function of extrinsic noise magnitude $CV$ for extrinsic noise in parameter $v_1$ with different correlation times, corresponding to a ratio of timescale separation $\lambda$ between intrinsic and extrinsic processes ranging from 6.4 to 25.6. The black cross denotes the result by Tostevin \textit{et al.}~\cite{Tostevin09} in the absence of extrinsic noise. The analytical solution of the power spectrum was calculated up to second order in $\epsilon$ (substitute $u = 1$ in Eqs.~\eqref{eq:result_Pe}, \eqref{eq:result_R}).}
\end{figure*}

In order to analyse the effects of extrinsic noise on the signal transduction process, we consider the simple two-stage gene expression model shown in Figure~\ref{fig:MI}(a), which contains mRNA molecules $M$ and protein molecules $A$, the concentrations of which fluctuate over time

\begin{equation}
\label{eq:MImodel}
\emptyset \overset{v_0}\rightarrow M \overset{d_0}\rightarrow \emptyset,  \,\,\,\,\,\,\,\,\,\,\,\,M\overset{v_1}\rightarrow M + A, \,\,\,\,\,\,\,\,\,\,\,\,A\overset{d_1}{\rightarrow}\emptyset.
\end{equation}
We again choose parameters for this motif such that they are consistent with those in the linear gene expression model. The values for $v_1$, $d_0$ and $d_1$ are the same as in Table~\ref{tab:Linearparameters}, while the mRNA production rate $v_0$ is calculated as
\begin{equation}
v_0=\phi_1^s v^*_0,
\end{equation}
where $\phi_1^s$ corresponds to the macroscopic steady-state concentration of the active promoter in the linear gene expression model and the asterisk refers to parameters as stated in Table~\ref{tab:Linearparameters}.

We are interested in how the rate of information transfer from mRNA to protein is affected by an extrinsic noise source that perturbs the translation process (extrinsic fluctuations in $v_1$). To quantify how well networks can transmit information in noisy environments, we can make use of the mutual information rate (MIR) as a metric. Calculation of the mutual information of trajectories is a challenging task \cite{Tostevin09, Bernardi:2015iq}, but for linear Gaussian processes an expression involving the power spectra of continuous-time input signal $s(t)$ and output signal $x(t)$ can be derived:
\begin{equation}
\label{eq:MI}
R(s(t),x(t)) = - \frac{1}{4\pi} \int_{-\infty}^{\infty}  \ln \bigg[ 1 - \frac{|P_{sx}(\omega)|^2}{P_{ss}(\omega)P_{xx}(\omega)} \bigg]   \diff{\omega}.
\end{equation}
Here, $P_{sx}(\omega)$ is the cross-power spectrum of $s(t)$ and $x(t)$ (the Fourier transform of the cross-correlation function of $s(t)$ and $x(t)$), and $P_{ss}(\omega)$ and $P_{xx}(\omega)$ are the power spectra of $s(t)$ and $x(t)$, respectively. Note that for non-Gaussian and/or non-linear systems this expression provides a lower bound for the channel capacity \cite{Gabbiani:1996vq}. We consider fluctuations in the mRNA concentration as the input signal, and fluctuations in the protein concentration as the output signal. In~\cite{Tostevin09}, Tostevin \textit{et al.} have derived the analytical expression for the MIR of the input and output trajectories of the motif in Figure~\ref{fig:MI}(a) in the absence of extrinsic noise. We extend this result by considering an external process that affects the translation process, causing the parameter $v_1$ to fluctuate over time. By substituting the analytical expressions for the (cross-)power spectra resulting from Eq.~\eqref{eq:P_as_sum} into Eq.~\eqref{eq:MI}, we are able to quantify how the accuracy of information transmission from mRNA to protein concentration is affected by this disrupting process.
\begin{table}
\caption{\label{tab:miparameters}Parameter values used for the information processing motif.}
\begin{ruledtabular}
\begin{tabular}{ll}
parameter & value ($s^{-1}$)  \\
\hline
$v_0$ & 0.000093   \\
$v_1$ & 0.028   \\
$d_0$ & 0.00019 \\
$d_1$ & 0.000064 \\
\end{tabular}
\end{ruledtabular}
\end{table}
Figure~\ref{fig:MI}(b) shows that in the absence of extrinsic noise, our approximate solution agrees with the solution by Tostevin \textit{et al.} (marked by the black cross)~\cite{Tostevin09}. The presence of extrinsic noise decreases the fidelity of information processing, and the extent of this effect depends on both the extrinsic noise magnitude $CV$ and the ratio of timescale separation $\lambda$, as defined in Eq.~\eqref{eq:lambda}, between the intrinsic and extrinsic processes. Extrinsic noise sources with long extrinsic correlation times (large $\lambda$) are less disruptive than those with shorter correlation times (smaller $\lambda$). This implies that slowly fluctuating environmental variables have a smaller negative effect on the MIR than external variables that fluctuate quickly. This result is intuitive, as in the extreme case where faster fluctuating extrinsic processes happen on roughly the same timescale as the intrinsic processes, it may prove harder to distinguish between signal and noise and information might be lost. In the limit of infinitely slowly fluctuating external variables, the value of the external variable is constant with respect to the intrinsic timescale, and will have no effect on the signal transduction process regardless of the noise magnitude.

\subsection{\label{sec:FFL}Robustness of feed-forward loop motifs}
\begin{figure*}
  \includegraphics[width=\textwidth]{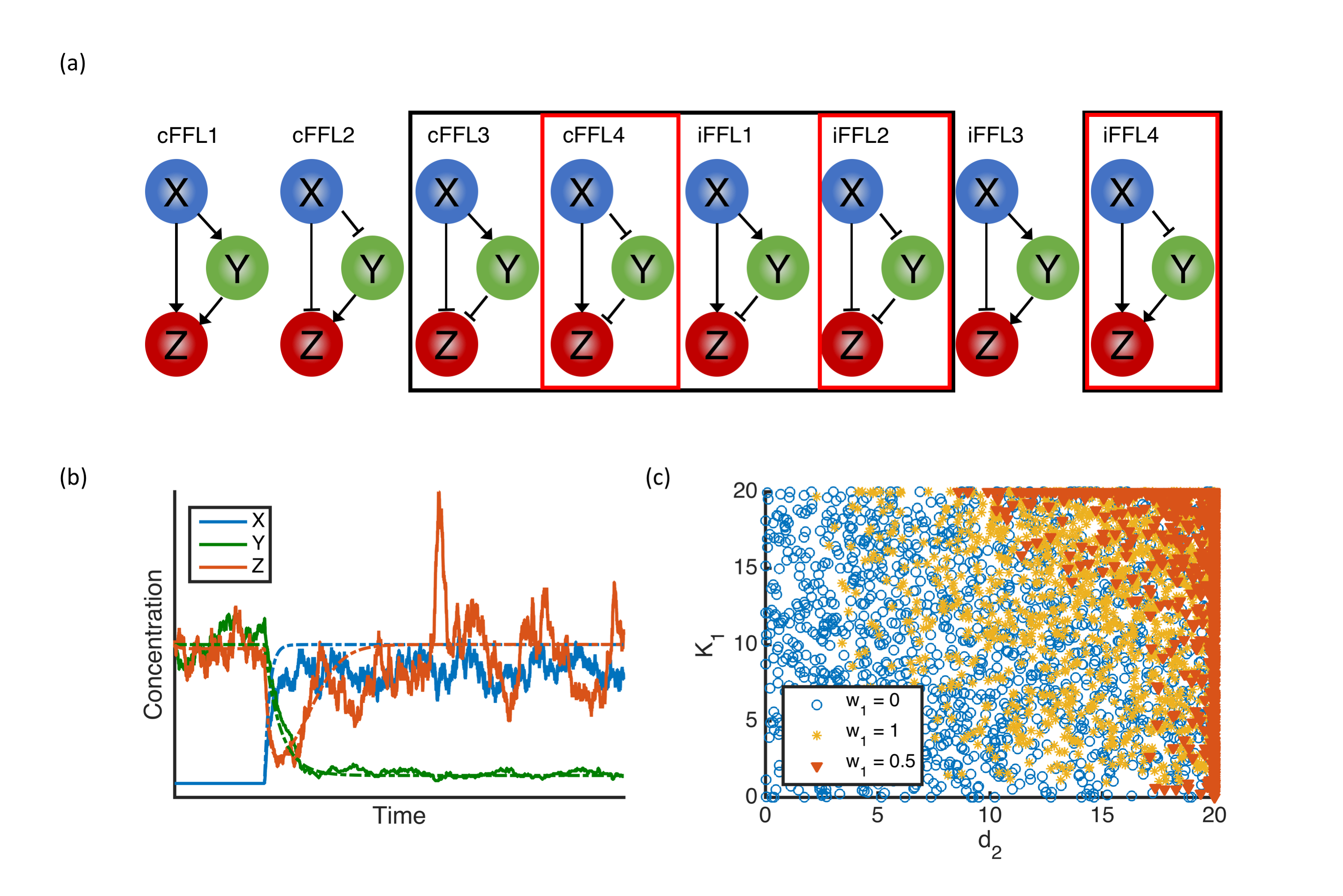}
  \caption{\label{fig:MOO}(a) All eight feed-forward loop topologies, where the arrow indicates the type of regulation (activation or repression). The red boxes denote the subset of optimal FFL topologies for the case of $\{ w_1 = 1, w_2 = 0\}$ in Eq.~\eqref{eq:score}, the black boxes denote the subset of optimal topologies when $\{ w_1 = 0.5, w_2 = 0.5\}$ (See Table~\ref{tab:fflfreq}). (b) Schematic of the FFL dynamic behaviour of our optimisation scheme. After the production of protein X is induced, this produced a drop in the concentration of protein Z. The concentration of Z then recovers to its original steady-state level. Solid lines denote example stochastic simulations of the concentration of proteins $X$ (blue), $Y$ (green), and $Z$ (red) over time, while dashed lines correspond to the respective deterministic dynamics. (c) How the parameter space of parameters $K_1$ and $d_2$ is constrained by considering an additional objective function for the system's stochastic behaviour. $w_1=0$ (blue circles) refers to the case where only the dynamic behaviour is prioritised,  $w_1=1$ (yellow asterisks) refers to the case where only the stochastic behaviour is prioritised, and $w_1=0.5$ (red downward-pointing triangles) refers to the case where both objectives are given equal importance.}
\end{figure*}

Feed-forward loops (FFLs) are capable of responding in a precise, robust manner to external signals~\cite{Mangan03}. These motifs are defined by a gene $X$ that regulates a second gene $Y$. Both $X$ and $Y$ then regulate a target gene $Z$. There are multiple types of FFLs (see Figure~\ref{fig:MOO}(a)) since regulation can take place either through activation or repression, and much effort has been devoted to extract the general features of each one. However, this is not straightforward because both the transient and equilibrium behaviour is characteristic of a particular system~\cite{Macia09}. For these reasons, constructing an optimal system that fulfils certain design requirements can be a considerable computational task. Here, we aim to present an efficient optimisation scheme to generate optimal parameters for a FFL to ensure it responds in a precise manner to input signals but remains robust to noise.
To do this, we devise two objective functions that quantify both the dynamic as well as the stochastic behaviour of the system. We aim to generate parameter sets for the network such that the system responds to a switching-on of the input signal $X$ by a negative pulse in the concentration of $Z$, before going back to its original steady state (red line, Figure~\ref{fig:MOO}(b)). Moreover, the variation around this steady state concentration of $Z$ for a noisy input signal $X$ should be minimal. Taking these requirements into account, the form of the objective function is
\begin{equation}
\label{eq:score}
S = w_1 c_{ODE} + w_2 c_{LNA},
\end{equation}
where $w_1$ and $w_2$ are the weights of the respective objective functions $c_{ODE}$ and $c_{LNA}$, with $w_1 + w_2 = 1$ and
\begin{align}
c_{ODE} &= \sum_{i = 1}^{5} q_i s_i , \label{eq:c_ode}\\
c_{LNA} &=  CV^{2}_{Z} = \frac{V(x_3^s)}{\average[]{x^s_3}^2}, \label{eq:c_lna}
\end{align}
where $x_3^s$ is the steady-state concentration of protein $Z$ and
\begin{align}
\label{eq:s1}
s_1 &= (\phi_1(t_{f})-\phi_1(t_{f}-\Delta t))^2,  \\
\label{eq:s2} 
s_2 &= (\phi_2(t_{f})-\phi_2(t_{f}-\Delta t))^2,\\
\label{eq:s3}
s_3 &= (\phi_3(t_{f})-\phi_3(t_{f}-\Delta t))^2,  \\
\label{eq:s4} 
s_4 &= (\phi_3(t_{f})-\phi_3(t_{0}))^2, \\
 \label{eq:s5}
s_5 &= \frac{\min(\phi_3)}{\average[]{\phi_3}}, 
\end{align}
with $\phi_i$ the macroscopic concentrations of the proteins $X$, $Y$, and $Z$ respectively, $q_i=\frac{1}{5}$, $i =1,\ldots, 5$ the subweights of each ODE (ordinary differential equation) objective $s_i$, $t_0=0$ the initial time point, $t_f=1000$ the final time point of the simulation, and $\Delta t=5$. To obtain $c_{ODE}$ we simulate the FFLs using an ODE solver, where Eqs.~\eqref{eq:s1}--\eqref{eq:s3} ensure each of the three system components reaches steady state, Eq.~\eqref{eq:s4} ensures that $Z$ reaches pre-input concentration, and Eq.~\eqref{eq:s5} aims to produce a significant drop in the concentration of $Z$ upon a change in input $X$ (Figure~\ref{fig:MOO}(b)). The score $c_{LNA}$ is obtained from our analytical solution (Eq.~\eqref{eq:tscale_stx} and Eq.~\eqref{eq:tscale_Vstx}). 

We used the most general model of the FFL from Mac\'\i a \textit{et al.} in~\cite{Macia09}, that is able to describe all eight FFL topologies. The macroscopic rate equations for this model are given by

\begin{equation}
\label{eq:FFLode}
\frac{\diff\mb{\phi}}{\diff t} = \begin{psmallmatrix} \alpha_0(1+\eta) - d_0 \phi_1 \\  \alpha_1 \bigg(\frac{1 + \beta_0 K_1 \phi_1}{1 + K_1 \phi_1} \bigg) - d_1 \phi_2 \\ \alpha_2\bigg(\frac{1 + \beta_1 K_2 \phi_1 + \beta_2 K_3 \phi_2 + \beta_3 K_2 K_3 \phi_1 \phi_2}{1 + K_2 \phi_1 + K_3 \phi_2 + K_2 K_3 \phi_1 \phi_2} \bigg) - d_2 \phi_3 \end{psmallmatrix}.
\end{equation}
In this model, $\alpha_i$ describes the basal production of the proteins $X$, $Y$, and $Z$, and $d_i$ denotes the degradation rate of a species. The type of regulatory interaction between the regulator gene and the gene it targets is described by parameters $\beta_i$, where values $\beta_i<1$ correspond to an inhibitory interaction as the production rate decreases proportional to the basal level, whereas $\beta_i > 1$ describes activation~\cite{Macia09}. $K_i$ describes the binding equilibrium of the regulator with the gene it targets. Extrinsic noise enters the model in the production rate of $X$, $\alpha_0$, with a magnitude of $CV = 0.5$. We fix the ratio of basal production/degradation ($\alpha_i/d_i = 100$) of all three molecular species to ensure that all species are sufficiently abundant for the LNA to hold. Initial conditions are $X(t_0)=0$, $Y(t_0)=Z(t_0)=100$. We generate $2\times 10^6$ random parameter sets for the 10 remaining system parameters and optimise for the objective score $S$ for the cases $\{ w_1 = 1, w_2 = 0\}$ and $\{ w_1 = 0.5, w_2 = 0.5\}$ (for details on the optimisation procedure, see Section~\ref{sec:Optimisation}). The $0.01\%$ top scoring parameter sets are then selected for either set of objective scores, and their corresponding topology is determined. The results in Table~\ref{tab:fflfreq} show that only three of eight possible topologies, shown in the red boxes in Figure~\ref{fig:MOO}(a), are present in the results when only the ODE objective function is considered ($w_1 = 1$), and that cFFL4 is the most prevalent motif. If both the ODE and LNA objectives are given equal importance ($w_1 = w_2 = 0.5$), then the cFFL3 and iFFL1 topologies are also present among the top-scoring parameter sets (black boxes, Figure~\ref{fig:MOO}(a)). However, since these do not occur when only considering the ODE criteria, this implies that they are not very suitable to give the desired dynamic behaviour. In addition, Mac\'\i a \textit{et al.} also find that the iFFL1 motif is not capable of producing the desired negative pulse~\cite{Macia09}. 

Since the cFFL4 topology is the most prevalent, this motif has the highest probability to produce the desired system behaviour. For this reason, we perform an optimisation within the local parameter space of the cFFL4 motif for $2\times 10^4$ randomly generated parameter sets and select the 1\,500 best-scoring sets. Figure~\ref{fig:MOO}(c) shows how the parameter space changes with the addition of the $c_{LNA}$ objective. If only the robustness of the network to noise is considered ($w_1=0$, blue circles), no specific parameter values for the parameters $K_1$ and $d_2$ are preferred. When only the dynamic behaviour of the FFL is prioritised ($w_1=1$, yellow asterisks), high values of $K_1$ and $d_2$ more likely result in the desired dynamics. Interestingly, when both the $c_{ODE}$ and $c_{LNA}$ objectives are given equal weight ($w_1=0.5$, red downward-pointing triangles) the parameter space is further constrained compared to optimising for a single objective. Thus, for robust FFLs the degradation rate of $Z$ and the binding equilibrium of protein $X$ to the gene that produces $Y$ needs to be tuned.

\begin{table}
\caption{\label{tab:fflfreq}Occurrence of FFL topologies of top (0.01\%) scoring parameter sets out of $2\times 10^6$ parameter sets.}
\begin{ruledtabular}
\begin{tabular}{lrr}
topology & $w_1 = 1$ & $w_1 = 0.5$  \\
\hline
iFFL2 & 1\% & 2\%  \\
cFFL4 & 75\% & 39.5\%   \\
iFFL4 & 24\% & 1\% \\
cFFL3 & 0\%  &  14.5\% \\
iFFL1 & 0\%  &  43\% \\
\end{tabular}
\end{ruledtabular}
\end{table}

\section{Discussion \& Conclusion}
In this work, we have derived an analytical framework to quantify the contribution of coloured extrinsic noise to fluctuations in gene expression. We have shown that when the conditions underlying the theory are satisfied, we are able to accurately describe the mean, variance and power spectrum of molecule number fluctuations subject to both intrinsic and extrinsic noise sources. Using several examples, we show that the theory is relevant in a wide range of applications, and can be used to distil the principles underlying fundamental system behaviour, noise sources, and information processing in biochemical networks.
 
Our framework relies on three main approximations: the linear-noise approximation, the separation between timescales of the intrinsic and extrinsic fluctuations, and the small extrinsic noise expansion. First, the LNA will give an accurate approximation of the CME when the molecular species populations are sufficiently large, when the nonlinearity in the reaction rates is sufficiently weak or else for special classes of biochemical systems (\cite{Grima15}; see discussion later). Second, our theory requires the extrinsic fluctuations to be slow with respect to the system's intrinsic dynamics. As most intrinsic processes happen on the timescale of seconds or sub-seconds and extrinsic fluctuations have a typical correlation time corresponding to the cell cycle period (many minutes), we expect that our timescale separation assumption is reasonable. The assumption of small extrinsic noise might appear at the first sight very limiting, however, in practice we find that the approximation yields sufficiently accurate results for noise magnitudes at least as large as $CV = 0.25$ for the mean protein number (Figure~\ref{fig:linearmodel}(b)) and variances (Figure~\ref{fig:linearmodel}(d)) of the linear model and $CV = 0.4$ for the means and variances of the regulated gene expression model (Figure~\ref{fig:negfeedback}(b)).
 
A more general issue with studying stochastic systems is computational speed. To obtain statistics of biochemical systems subject to both intrinsic and extrinsic noise with reasonable confidence levels, one needs to simulate many trajectories of the system for a considerable time. The advantage of obtaining closed-form expressions for these statistics is that parameter values simply need to be substituted, and there is no need to re-evaluate the system. For example, it takes approximately 3 hours to generate one data point (100 trajectories of $10^8$ seconds each) for the model in Figure~\ref{fig:linearmodel}(a), whereas evaluating the analytical expressions for the mean and variance of each molecular component takes less than a second on a typical desktop computer. As pointed out in Section~\ref{sec:automated}, the limiting step in the automated sum evaluation is the calculation of stationary state $\mb\phi^s$. This task becomes increasingly computationally intensive for more complex systems. For this reason, the limit of complexity that we can study is determined by the computational power available. For specific cases, a more specialised external solver could be employed to accelerate this task.
 
Due to this speed-up, we have been able to use the analytical framework to perform a computationally efficient multi-objective optimisation of FFLs. With this optimisation routine, we are able to explore both network topology and parameter space to generate systems with optimal dynamic and stochastic features, which is generally infeasible for non-trivial systems using simulation-based approaches. Our analysis shows that even in simple networks such as FFLs, there exists a complex relationship between system structure and function. With this optimisation scheme we are able to quickly generate recommendations for an optimal network topology and parameter ranges. Compared to optimisation using stochastic simulation algorithms, this optimisation scheme gives an improvement in computational time of several orders of magnitude. The results of the optimisation scheme suggest that not every FFL motif is capable of producing a specific dynamic response, and that not all FFL types have the same extrinsic noise tolerance. Although there does not appear to be a trade-off between these two objectives, choosing optimal networks and their parameters such that they fulfil both requirements can be a substantial task given the high dimensionality of the problem. In this case, combining deterministic dynamics with stochastic analysis of equilibrium behaviour is an efficient and effective approach.
 
Since analytical expressions for both intrinsic and extrinsic contributions to variability can be obtained, the proposed method allows a systematic analysis of how changing the properties of extrinsic fluctuations affects intrinsic variability and total noise in gene expression models. Similarly, these expressions can provide predictions on which network parameters are susceptible to perturbations and contribute to high variability and can be used as a tool for stochastic sensitivity analysis. Such a method can be of interest for synthetic biology applications as it could provide universal design principles for network construction that exploits (suppresses) the positive (negative) effects of cellular stochasticity.
 
The framework developed here rests on the validity of the linear-noise approximation first and foremostly. This limits the current approach to analysis of nonlinear biochemical systems with large numbers of molecules in all species or else to those systems with arbitrary number of molecules but weakly nonlinear reaction rates.  However, we note that the linear-noise approximation has been, over the past decade, extended to estimate the first and second moments of the molecule number distributions of nonlinear biochemical systems in which one ore more molecular species is present in low copy numbers (\cite{Frohlich:2016ij, Grima:2010jl, grima2011accurate, grima2012study}). The corrections to the LNA power spectrum of fluctuations due to low molecule numbers have also been systematically studied \cite{Thomas:2013em,thomas2014system}. Hence by starting from these frameworks and repeating the same analysis as we performed here, i.e., applying the assumption of timescale separation between intrinsic and extrinsic noise and subsequently assuming small extrinsic noise, would likely result in a new theory which overcomes the major limitations of the present approach.
 
In conclusion, we have proposed a fast, systematic analytical framework to assess the effects of coloured environmental noise on biochemical systems. We have shown that the mathematical framework provides accurate predictions of system characteristics for a wide range of biological networks. Given the speed and flexibility of our approach, the research community can now further access the sources of variability in gene expression data. This will lead to a better understanding of how biological systems exploit or suppress environmental signals. There is, thus, the potential to uncover new design principles to aid the construction of new, robust \textit{in vivo} synthetic circuits.

\begin{acknowledgments}
E.K and C.F. were supported by HFSP Research grant RGP0025/2013. R. G. was supported by BBSRC grant BB/M025551/1. C.F. acknowledges fruitful discussions with Benjamin Lindner. 
\end{acknowledgments}
\clearpage
\begin{widetext}
\appendix

\section{\label{sec:noise_construction}Construction of the noise variables}

To introduce extrinsic fluctuations to a rate constant $c_k$, we
multiply it with a lognormal variable $\bar\nu_k(t)$,
\begin{align}
  \bar{c}_k(t) &= {c}_k \bar\nu_k(t) = c_k \left(1 + \bar\eta_k(t)\right) .
\end{align}
We require $\average{\bar{c}_k(t)} = c_k$ and also define a shifted stochastic
variable $\bar\eta_k(t)$ that will be needed in the small noise expansion.
The lognormal variables $\bar\nu_k(t)$ can be constructed from normally distributed
variables $\bar\mu_k(t)$
with variances $\average{\bar\mu_k\,\bar\mu_k} = \epsilon_k^2$ and inverse correlation
times $K_k$.
The latter may be described by the Ornstein-Uhlenbeck process
\cite{Gardiner1990}
\begin{align}
  \label{eq:def_mu}
  \diff \bar\mu_k(t) = - K_k \bar\mu_k(t) \diff t + \sqrt{2 K_k} \epsilon_k \diff W(t) .
\end{align}
We define the lognormally distributed stochastic variable by
\begin{align}
  \label{eq:def_nu}
  \bar\nu_{k}(t) = \exp\left(\bar\mu_{k}(t) - \frac{1}{2} \epsilon_{k}^2\right)
\end{align}
and use Wick's theorem~\cite{Bellac1991} to calculate its mean
(only even powers in $\bar\mu_k^c$ do not vanish):
\begin{align}
  \average{\bar\nu_{k}} &= \exp\left(-\frac{1}{2} \epsilon_{k}^2\right)
  \sum_{c=0}^{\infty} \frac{1}{(2 c)!} \left\langle \bar\mu_{k}^{2 c} \right\rangle
  = \exp\left(-\frac{1}{2} \epsilon_{k}^2\right)
  \sum_{c=0}^{\infty} \frac{1}{2^c c!}
  {\underbrace{\average{\bar\mu_k\,\bar\mu_k}}_{\epsilon_{k}^2}}^c = 1
\end{align}
in agreement with the requirement $\average{\bar{c}_k(t)} = c_k$.
The shifted lognormal stochastic variable is then
\begin{align}
  \label{eq:def_eta}
  \bar\eta_{k}(t) = \exp\left(\bar\mu_{k}(t) - \frac{1}{2} \epsilon_k^2\right) - 1 .
\end{align}

\section{\label{sec:lognormal_corr}Arbitrary time correlation functions of lognormal stochastic variables}
Given a set of independent normally distributed stochastic variables
$\{\bar\mu_k\}$, a tuple of indices $(r_1, \dots, r_{n})$
and a tuple of times $(t_1, \ldots, t_{n})$
we define $\mu_k \equiv \bar\mu_{r_k}(t_k)$,
and the two-point time correlation functions
$\Delta_{ij}$ for $1 \leq i < j \leq n$,
\begin{align}
  \Delta_{ij} \equiv \average{\mu_i \mu_j}
\end{align}
for all $i < j$.
We consider a smooth function and its derivatives
\begin{align}
  \label{eq:Y}
  y_{k} &\equiv V(\mu_{k}) = \sum_{c=0}^{\infty} \frac{a_{c}}{c!} \mu_{k}^{c}
  \,, &
  y_{k}^{(m)}(\mu_{k}) &\equiv \frac{\diff^{m} V(\mu_{k})}{\diff \mu_{k}^{m}}
  = \sum_{c=m}^{\infty} \frac{1}{(c-m)!} a_{c} \, \mu_{k}^{c-m}
\end{align}
to derive the generalisation of the $n=2$ result by
Malakhov~\cite{Malakhov1978} for the $n$-point time correlation function
\begin{align}
  \label{eq:y_corr}
  \average{y_{1} \dots y_{n}} &=
  \sum_{c_{1} = 0}^{\infty} \dots \sum_{c_{n} = 0}^{\infty}
  \frac{a_{c_1} \dots a_{c_n}}{c_{1}! \dots c_{n}!}
  \average{\mu_{1}^{c_1} \dots \mu_{n}^{c_n}} .
\end{align}
According to Wick's theorem~\cite{Bellac1991}, the correlation functions
$\average{\mu_{1}^{c_1} \dots \mu_{n}^{c_n}}$
decompose into sums of partitions into two-point correlation functions and
they are zero for odd $n$.  We apply the theorem partially to isolate the
two-point correlation functions $\Delta_{ij} = \average{\mu_{i} \mu_{j}}$
with $i < j$,
\begin{align}
  \label{eq:y_corr1}
  \average{y_{1} \dots y_{n}} &=
  \sum_{l_{1} = 0}^{\infty} \dots \sum_{l_{n} = 0}^{\infty}
  \sum_{u = 0}^{\infty}
  \hspace{-.7em}\sum_{\hspace{.7em}|\bd^n| = u}
  \Biggl(
    \frac{c_{1}! \dots c_{n}!}{l_{1}! \dots l_{n}! \prod\limits_{i<j} d_{ij}!}
    \frac{a_{c_1} \dots a_{c_n}}{c_{1}! \dots c_{n}!}
    \bigl\langle\mu_{1}^{l_1}\bigr\rangle \dots \bigl\langle\mu_{n}^{l_n}\bigr\rangle
    \prod_{
      \substack{ i, j = 1 \\ i < j }
    }^{n}
    \Delta_{ij}^{d_{ij}}
  \Biggr)
\end{align}
where the inner sum is taken over all tuples
$\bd^n = (d_{12}, d_{13}, d_{23}, \dots, d_{(n-1)\,n})$
with $|\bd^n| = \sum d_{ij} = u$.
There are $c_{k}!/l_{k}!$ possibilities to assign $m_{k} = c_{k}-l_{k}$
from a total of $c_{k}$ factors $\mu_k$ to the $u$ $\Delta_{ij}$ pairs.
However, to obtain the number of \emph{different} partitions into the pairs, the
product of the former must be divided by the product of $d_{ij}!$ permutations
of $d_{ij}$ identical factors $\Delta_{ij}$.
We notice that
$l_{k} = c_{k} - m_{k}$ to recognise the derivatives $y_{k}^{(m_k)}$
from Eq.~\eqref{eq:Y} so finally
\begin{align}
  \label{eq:timecorr_nu_general}
  \average{y_{1} \dots y_{n}} &=
  \sum_{u = 0}^{\infty}
  \hspace{-.7em}\sum_{\hspace{.7em}|\bd^n| = u}
  \Biggl(
    \prod_{k=1}^{n} \average{y_{k}^{(m_k)}(\mu_{k})}
    \prod_{
      \substack{ i, j = 1 \\ i < j }
    }^{n}
    \frac{\Delta_{ij}^{d_{ij}}}{d_{ij}!}
  \Biggr) .
\end{align}
For $n=2$ with $k = m_{1} = m_{2} = d_{12}$ and
$B_{\mu}[\tau] = \Delta_{12}(\tau) = \average{\mu(0)\mu(\tau)}$
we recover the result by Malakhov~\cite{Malakhov1978},
\begin{align}
  \label{eq:malakhov}
  B_{y}[\tau] = \average{y(0)y(\tau)} - \average{y}^2
  = \sum_{k=1}^{\infty} \frac{1}{k!}
  \bigl\langle y^{(k)}(\mu)\bigr\rangle^2 B_{\mu}^{k}[\tau] .
\end{align}
\\

\paragraph{Evaluation for normal stochastic variables}
With normally distributed $y_k = \mu_k$ (mean 0) the term
$\langle{}y_{k}^{(m_k)}(\mu_{k})\rangle$ is 1 for $m_k = 1$ and 0 else.
Consequently, $m_k = 1$ for $k \in \{1,\dots,n\}$, that is each index must
occur exactly once in the product of two-point correlation functions in
Eq.~\eqref{eq:timecorr_nu_general} so also all $d_{ij}$ are 1 and Wick's theorem
is recovered.
\\

\paragraph{Evaluation for lognormal stochastic variables}
Lognormally distributed $y_k = \nu_k$ with mean 1 is invariant under
differentiation with respect to $\mu_k$ (see Eq.~\ref{eq:def_nu}) so the term
$\langle{}y_{k}^{(m_k)}(\mu_{k})\rangle$ becomes identical 1 for all $m_k$.
This leads to significant simplification of Eq.~\eqref{eq:timecorr_nu_general}
and we use the multinomial theorem to obtain
\begin{align}
  \label{eq:timecorr_nu}
  \average{\nu_1 \ldots \nu_n}
  = \sum_{u = 0}^{\infty}
  \hspace{-.7em}\sum_{\hspace{.7em}|\bd^n| = u}
  \Biggl(\,
    \prod_{
      \substack{ i, j = 1 \\ i < j }
    }^{n}
    \frac{\Delta_{ij}^{d_{ij}}}{d_{ij}!}
  \Biggr)
  = \sum_{k=0}^{\infty} \frac{1}{k!}
  \Biggl(
    \sum_{
      \substack{ i, j = 1 \\ i < j }
    }^{n}
    \Delta_{ij}
  \Biggr)^{k}
  = \exp\Biggl(
    \sum_{
      \substack{ i, j = 1 \\ i < j }
    }^{n}
    \Delta_{ij}
  \Biggr) .
\end{align}
\\

\paragraph{Evaluation for shifted lognormal stochastic variables}
The mean of the stochastic variables $\eta_k$ in Eq.~\eqref{eq:def_eta} is 0
and all derivatives with respect to $\mu_k$ are identical to $\nu_k$ in
Eq.~\eqref{eq:def_nu} with mean 1.
Therefore, the product of $\langle{}y_{k}^{(m_k)}(\mu_{k})\rangle$ terms
in Eq.~\eqref{eq:timecorr_nu_general} vanishes if $m_k = 0$ for any
$k \in \{1,\dots,n\}$ and is 1 else.
The final result is
\begin{align}
  \average{\eta_1 \ldots \eta_n}
  = \sum_{u = 0}^{\infty}
  \hspace{-.7em}\sum_{\hspace{.7em}|\bd^n| = u}\raisebox{3pt}{\hspace{-.9em}$'$}\hspace{0.5em}
  \Biggl(
    \prod_{
      \substack{ i, j = 1 \\ i < j }
    }^{n}
    \frac{\Delta_{ij}^{d_{ij}}}{d_{ij}!}
  \Biggr)
\end{align}
where the prime denotes the condition that for each $j \in \{1, \dots, n\}$
there is a $i < j$ or $k > j$ such that $d_{ij} \neq 0$ or $d_{jk} \neq 0$
(consequently $m_j \neq 0$).
For example, with $n = 2$ we obtain
$\average{\eta_1 \eta_2} = \exp\left(\Delta_{12}\right) - 1$.
Evaluation for $r_1 = r_2 = k$ at equal times gives the variance
$\average{\bar\eta_k^2} = \exp\left(\epsilon_k^2\right) - 1$
for $\bar\eta_k$ in Eq.~\eqref{eq:def_eta}
where the variance of $\mu_k$ is $\epsilon_k^2 = \average{\mu_k\,\mu_k}$.

\section{\label{sec:small_noise_xs}Small noise expansion for the mean}
The calculation of the mean concentrations $\average{\mathbf x^s}$
in Eq.~\eqref{eq:small_noise_mean} is mediated by the multi-index notation 
defined by Eq.~\eqref{eq:def:multi_r} so the correlation functions
in the small noise expansion read
$\average{\eta_{r_1} \ldots \eta_{r_n}}$
and can be evaluated by means of Eq.~\eqref{eq:timecor_eta}:
\begin{align}
  \average{\mathbf x^s} = \sum_{n=0}^{\infty} \sum_{\# \br = n}
  \mb\phi^s_{(\br)}
  \sum_{u = \lfloor\frac{n + 1}{2}\rfloor}^{\infty}
  \hspace{-.7em}\sum_{\hspace{.7em}|\bd^n| = u}\raisebox{3pt}{\hspace{-.9em}$'$}\hspace{0.5em}
  ~\prod_{
    \substack{ i, j = 1 \\ i < j }
  }^{n}
  \frac{1}{d_{ij}!}
  \Bigl(\Gamma_{r_i r_j}\Bigr)^{d_{ij}} .
\end{align}
$\Gamma_{i j}$ denotes the covariances
$\average{\mu_{i}(t)\,\mu_{j}(t)}$ of the normal stochastic variables.
We rearrange the order of summation with $u$ as the principal summation index,
then $n$ runs from $0$ to $2 u$, and obtain the final result in
Eq.~\eqref{eq:result_mean}.

\section{\label{sec:small_noise_var}Small noise expansion for the covariance matrix}
The small noise expansion of $\mathbf V\left(\mb\xi\right)$ in analogy to
$\average{\mathbf x^s}$ is detailed in the main text, see
Eq.~\eqref{eq:result_Vxi}.  For $\mathbf V(\mb\phi^s)$ we use the
multi-index $\bq = (q_1, q_2)$, $|\bq| = q_1 + q_2$,
and the Taylor series of $\mb\phi^s$ in Eq.~\eqref{eq:small_noise_mean}
to expand Eq.~\eqref{eq:def:variances}:
\begin{align}
  \label{eq:small_noise_Vstphi}
  \mathbf V(\mb\phi^s) =
  \sum\limits_{n = 0}^{\infty} \sum\limits_{|\bq| = n}
  \hspace{-1.2em}\sum\limits_{\hspace{1.2em}\# \br^1 = q_1}
  \hspace{-1.2em}\sum\limits_{\hspace{1.2em}\# \br^2 = q_2}
  \mb\phi^s_{(\br^1)} \mb\phi^{s\,T}_{(\br^2)}
  \left(
    \average{\eta_{(\br^1)} \eta_{(\br^2)}}
    - \average{\eta_{(\br^1)}} \average{\eta_{(\br^2)}}
  \right) .
\end{align}
To evaluate the correlation functions
we use Eq.~\eqref{eq:timecor_eta}
and the index functions $f^{2}_{i j}$
from Eq.~\eqref{eq:delta2} to obtain
\begin{align}
  \label{eq:stphi_corr_eval_1}
  \average{
    \eta_{(\br^1)} \eta_{(\br^2)}
  }
  &= \sum_{u = \lfloor\frac{n + 1}{2}\rfloor}^{\infty}
  \hspace{-.7em}\sum_{\hspace{.7em}|\bd^n| = u}\raisebox{3pt}{\hspace{-.9em}$'$}\hspace{0.5em}
  ~\prod_{
    \substack{ i, j = 1 \\ i < j }
  }^{n}
  \frac{1}{d_{ij}!}
  \Bigl(\Gamma_{f^2_i f^2_j}\Bigr)^{d_{ij}} ,
\end{align}
\begin{align}
  \label{eq:stphi_corr_eval_2}
  \average{ \eta_{(\br^1)} } \average{ \eta_{(\br^2)} }
  &= \sum_{v = \lfloor\frac{n_{1} + 1}{2}\rfloor}^{\infty}
  \hspace{-.7em}\sum_{\hspace{.7em}|\bd^{n_{1}}| = v}\raisebox{3pt}{\hspace{-.9em}$'$}\hspace{0.5em}
  ~\prod_{
    \substack{ i, j = 1 \\ i < j }
  }^{n_{1}}
  \frac{1}{d_{ij}!}
  \Bigl(\Gamma_{r^1_i r^1_j}\Bigr)^{d_{ij}}
  \quad\times
  \sum_{w = \lfloor\frac{n_{2} + 1}{2}\rfloor}^{\infty}
  \hspace{-.7em}\sum_{\hspace{.7em}|\bd^{n_{2}}| = w}\raisebox{3pt}{\hspace{-.9em}$'$}\hspace{0.5em}
  ~\prod_{
    \substack{ i, j = 1 \\ i < j }
  }^{n_{2}}
  \frac{1}{d_{ij}!}
  \Bigl(\Gamma_{r^2_i r^2_j}\Bigr)^{d_{ij}} .
\end{align}
Each term in Eq.~\eqref{eq:stphi_corr_eval_2} appears as well in
Eq.~\eqref{eq:stphi_corr_eval_1} when $u = v + w$.
The other way round, every term in \eqref{eq:stphi_corr_eval_1}
that contains only two-point
correlation functions that occur in one of the two factors in
Eq.~\eqref{eq:stphi_corr_eval_2} cancels in the difference
\begin{align}
  \label{eq:var_difference_final}
  \average{\eta_{(\br^1)} \eta_{(\br^2)}}
  - \average{\eta_{(\br^1)}} \average{\eta_{(\br^2)}}
  =
  \sum_{u = \lfloor\frac{n + 1}{2}\rfloor}^{\infty}
  \hspace{-.7em}\sum_{\hspace{.7em}|\bd^n| = u}\raisebox{3pt}{\hspace{-.9em}$''$}\hspace{0.5em}
  ~\prod_{
    \substack{ i, j = 1 \\ i < j }
  }^{n}
  \frac{1}{d_{ij}!}
  \Bigl(\Gamma_{f^2_i f^2_j}\Bigr)^{d_{ij}}
  \,.
\end{align}
The new restriction indicated by the second prime in the sum,
there is at least one $d_{ij} \neq 0$ for which $i \le n_1$ and $j > n_1$,
asserts that only those terms from Eq.~\eqref{eq:stphi_corr_eval_1} are
kept that do not cancel with the corresponding term of the sum in
Eq.~\eqref{eq:stphi_corr_eval_2}.  We substitute this result into
Eq.~\eqref{eq:small_noise_Vstphi} and change the order of summation.
For $n < 2$ this restriction cannot be fulfilled so the sum
in the result Eq.~\eqref{eq:result_Vphis} in the main text
starts with $u = 1$ and $n = 2$.

\section{\label{sec:small_noise_Pe}First integral for the spectrum matrix}

To compute Eq.~\eqref{eq:integrals_Pe} for the spectrum matrix $\bP_e(\omega)$
we first evaluate the two-point time correlation functions
$\Delta_{ij}(t_i - t_j)$ of the normal
stochastic variables  $\mu_{i}$ according to Eq.~\eqref{eq:delta},
\begin{multline}
  \frac{1}{2\pi} \int_{0}^{\infty} e^{- \ii \omega \tau}
  \prod_{
    \substack{ i, j = 1 \\ i < j }
  }^{n}
  \frac{1}{d_{ij}!}
  \Bigl(\Delta_{f^2_i f^2_j}(t_i - t_j)\Bigr)^{d_{ij}}
  \dtau
  =
  \frac{1}{2\pi} \int_{0}^{\infty} e^{- \ii \omega \tau}
  \prod_{
    \substack{ i, j = 1 \\ i < j }
  }^{n}
  \frac{1}{d_{ij}!}
  \left(\Gamma_{f^2_i f^2_j} e^{- K_{f^2_i} | t_i - t_j |}\right)^{d_{ij}}
  \dtau
  \\
  =
  \frac{1}{2\pi}
  \int_{0}^{\infty} e^{- \bigl(
      \ii \omega +
      \hspace{-1em}
      \sum\limits_{\hspace{1em}
	i < j = 1
      }^{n}
      d_{ij} K_{f^2_i} | t_i - t_j | \tau^{-1}
    \bigr)\tau}
  \dtau
  \prod_{
    \substack{ i, j = 1 \\ i < j }
  }^{n}
  \frac{1}{d_{ij}!} \left(\Gamma_{f^2_i f^2_j}\right)^{d_{ij}} .
\end{multline}
So far we have not treated the times $t_i$, $t_j$ precisely and need to follow
them back to equation Eq.~\eqref{eq:def_Pe} where $t_1 = \tau$ corresponds to
$\mb\eta^1$ and $t_2 = 0$ to $\mb\eta^2$.
The information needed for their correct evaluation is traceable in the definition
of the index function $f^2_i$ in Eq.~\eqref{eq:delta2} that maps to components of $\mb\eta^1$ if
$i \le \# \br^1$ ($t_i = \tau$) and to $\mb\eta^2$ else ($t_i = 0$).
The difference $|t_i - t_j| \tau^{-1}$ becomes either $0$ or $1$ and the sum in
the exponent reduces to pairs $i,j$ that obey $1 \le i \le \# \br^1 < j \le n$.
With finite and positive inverse correlation times $K_i$, the exponent has a
negative real part so the integral with the solution
\begin{align}
  \frac{1}{2\pi}
  \Biggl(
    \ii \omega +
    \sum_{i=1}^{\# \br^1}
    \hspace{-2em}\sum_{\hspace{2em}j = \# \br^1 + 1}^{n}
    d_{ij} \, K_{f^2_i}
  \Biggr)^{-1}
\end{align}
exists.
Finally, we evaluate the index function $f^2_i = r^1_i$ for $i\le\#\br^1$,
denote the double sum as $\Theta(\bd^n, \br^1)$ in
Eq.~\eqref{eq:integrals_Pe_final} and add the complex conjugate of the whole
expression to obtain $\bP_e(\omega)$ in Eq.~\eqref{eq:result_Pe}.

\section{\label{sec:small_noise_Pi}Second integral for the spectrum matrix}

To calculate the spectrum matrix
$\average{\bP_{i}(\omega)}$ in Eq.~\eqref{eq:def_Pi}, we expand
Eq.~\eqref{eq:def_R} with the Taylor coefficients for $\bA(\mb\eta_1)$,
using Eq.~\eqref{eq:def:multi_r}, and $\bC(\mb\eta_1, \mb\eta_2)$,
using Eq.~\eqref{eq:def:multi_s} and \eqref{eq:taylor_C_s},
\begin{align}
  \bR(\omega)
  &= \int_{0}^{\infty} e^{-\left(-\bA^0 + i\omega\right)\tau}
  \average[]{\sum_{c=0}^{\infty} \frac{\tau^c}{c!} \Biggl(
      \sum_{q=1}^{\infty}
      \hspace{-0.5em}\sum\limits_{\hspace{0.5em}\# \br = q}
      \bA_{(\br)} \eta_{(\br)}
    \Biggr)^c
    \Biggl( \sum_{a=0}^{\infty}
      \hspace{-0.5em}\sum\limits_{\hspace{0.5em}\# \br = a}
      \hspace{-0.5em}\sum\limits_{\hspace{0.5em}\# \bs = a}
      \bC_{(\br,\bs)} \eta_{(\br,\bs)}
    \Biggr)
  } \dtau .
\end{align}
The term in the average is a sum of terms of the form
\begin{align}
  \frac{\tau^c}{c!}
  \Biggl(
    \hspace{-1.2em}\sum\limits_{\hspace{1.2em}\# \br^1 = q_1}
    \bA_{(\br^1)} \eta_{(\br^1)}\Biggr)
  \ldots
  \Biggl(
    \hspace{-1.2em}\sum\limits_{\hspace{1.2em}\# \br^c = q_c}
    \bA_{(\br^c)} \eta_{(\br^c)}\Biggr)
  \Biggl(
    \hspace{-1.9em}\sum\limits_{\hspace{1.9em}\# \br^{c+1} = a}
    \hspace{-0.5em}\sum\limits_{\hspace{0.5em}\# \bs = a}
    \bC_{(\br^{c+1},\bs)} \eta_{(\br^{c+1},\bs)}
  \Biggr).
\end{align}
With the multi-index $\bq = (q_1, \ldots, q_c)$ from
Eq.~\eqref{eq:def:multi_q} we change to $n = |\bq| + a$
(the order in $\mb\eta$) as principal sum index,
\begin{multline}
  \label{eq:R_multiindex}
  \bR(\omega) =
    \sum\limits_{n=0}^{\infty}
    \sum\limits_{a=0}^{n}
    \hspace{-1.4em}\sum\limits_{\hspace{1.4em}|\bq| = n - a}
    \hspace{-1.2em}\sum\limits_{\hspace{1.2em}\# \br^1 = q_1}
    \hspace{-0.5em}\ldots
    \hspace{-1.2em}\sum\limits_{\hspace{1.2em}\# \br^c = q_c}
    \hspace{-0.5em}\sum\limits_{\hspace{0.5em}\# \bs = a}
  \\\nonumber\times
  \int_{0}^{\infty} e^{-\left(-\bA^0 + i\omega\right)\tau} \, \frac{\tau^c}{c!}
  \Bigl(
    \bA_{(\br^1)} \ldots \bA_{(\br^c)} \bC_{(\br^{c+1},\bs)}
    \average[]{\eta_{(\br^1)} \ldots \eta_{(\br^c)} \eta_{(\br^{c+1},\bs)}}
  \Bigr) \dtau .
\end{multline}
The sum $\sum_{|\bq| = n - a}$ is carried out over all possible $c$,
$(q_1, \dots, q_c)$ with $q_1 + \dots + q_c +a = n$.  The correlation
functions are calculated according to Eq.~\eqref{eq:timecor_eta}.
After changing the order of summation, a comparison to $\bR(\omega)$ in
Eq.~\eqref{eq:result_R} gives
\begin{multline}
  \Bigl(
    - \bA\!^0 + \theta(\bd^c, |\bq|, \br^{c+1}, \bs) + \ii \omega
    \Bigr)^{-(c + 1)}
  =
  \frac{1}{c!}
  \int_{0}^{\infty} e^{-\bigl(
      -\bA^0 + \ii \omega +
      \hspace{-1em}
      \sum\limits_{\hspace{1em}
	i < j = 1
      }^{n}
      d_{ij} K_{f^{c+1}_i} | t_i - t_j | \tau^{-1}
    \bigr)\tau}
  \tau^c \dtau
  \\ =
  \Bigl(
    -\bA^0 + \ii \omega +
    \hspace{-1em}
    \sum\limits_{\hspace{1em}
      \substack{ i, j = 1 \\ i < j }
    }^{n}
    d_{ij} K_{f^{c+1}_i} | t_i - t_j | \tau^{-1}
    \Bigr)^{- (c + 1)}
\end{multline}
where the sum in the exponent stems from $\Delta_{ij}(t_i - t_j)$ in
Eq.~\eqref{eq:delta} and
\(
  \int_{0}^{\infty} e^{- a \tau} \, \tau^c \dtau
  = \left(- \frac{\diff}{\diff a}\right)^c \int_{0}^{\infty} e^{- a \tau} \dtau
  = \frac{c!}{a^{c + 1}}
\)
was used in the last equality.
We identify the $\theta$ function as the sum in the last term that we evaluate
following the arguments in section~\ref{sec:small_noise_Pe}.
The difference $|t_i - t_j| \tau^{-1}$ is $0$ or $1$ and non zero if and only if
$f^{c+1}_j$ in Eq.~\eqref{eq:delta2} maps to a component of $r^{c+1}$, that is
$j > |\bq|$, and the index function $f^{c+1}_i$ either maps to a component of
$\br^{i}$, $i \in \{1,\dots,c\}$ (corresponding to $t_1$ in Eq.~\ref{eq:def_R})
nd $\sigma_{i-|\bq|} = 2$ (corresponding to $t_2$) or also $i > |\bq|$ and
$\sigma_{i-|\bq|} \neq \sigma_{j-|\bq|}$ (so $t_1 \neq t_2$).
This result is formalised in Eq.~\eqref{eq:integrals_Pi_final} and \eqref{eq:Pi_beta_gamma} in the main text.
For simplification of the notation in Eq.~\eqref{eq:integrals_Pi_final},
$K_{f^{c+1}_i}$ is evaluated to $K_{r^{c+1}_{j-|\bq|}}$ which is allowed due to the
$\Gamma_{f^{c+1}_i f^{c+1}_j}$ in Eq.~\eqref{eq:result_R} that is proportional
to $\delta_{f^{c+1}_i f^{c+1}_j}$.

\section{\label{sec:theory_example}Exemplary evaluation of the small noise expansion}

While the closed-form expressions obtained from the small noise expansion are
well suited for automated evaluation, the notation is rather complicated and
will be unfamiliar to most readers.
To facilitate reading of the sums to the reader we evaluate the
first terms in more detail here.
The simplest of the sums is Eq.~\eqref{eq:result_mean} for the mean
concentrations
\begin{align*}
  \hspace{-0.5em}
  \average{\mathbf x^s} =
  \sum_{u = 0}^{\infty}
  \sum_{n=0}^{2 u} \sum_{\# \br = n}
  \mb\phi^s_{(\br)}
  \hspace{-.7em}\sum_{\hspace{.7em}|\bd^n| = u}\raisebox{3pt}{\hspace{-.9em}$'$}\hspace{0.5em}
  \,\prod_{
    \substack{ i, j = 1 \\ i < j }
  }^{n}
  \frac{1}{d_{ij}!}
  \Bigl(\Gamma_{r_i r_j}\Bigr)^{d_{ij}}
  \hspace{-0.5em}
\end{align*}
for which we evaluate all terms for $u=0$ and $u=1$:

\hspace{.05\textwidth}
\begin{minipage}{.92\textwidth}
  \bigskip
  \begin{tabular}{p{1.5cm}p{1.5cm}p{2.0cm}l}
    $u=0$ & $n=0$ & $\br=()$ & \(
    \mb\phi^s_{(\br)} = \mb\phi^s_{()} = \mb\phi^s(\vect{0})
    \)\\
  \end{tabular}
  \smallskip

  \begin{tabular}{p{1.5cm}p{1.5cm}p{2.0cm}l}
    $u=1$ & $n=0$ & $\br=()$ & 0 \\
	  & $n=1$ & $\br=(i)$ & 0 \\
	  & $n=2$ & $\br=(i,j)$ & \(
	  \mb\phi^s_{(i,j)} \Gamma_{ij}
	  = \frac{1}{2} \pfrac{\eta_i} \pfrac{\eta_j}
	  \left.\mb\phi^s(\mb\eta)\right|_{\mb\eta = \vect{0}} \delta_{ij} \epsilon_i^2
	  = \frac{1}{2} \pfrac[\unskip^2]{\eta_i^2}
	  \left.\mb\phi^s(\mb\eta)\right|_{\mb\eta = \vect{0}} \epsilon_i^2
	  \)\\
  \end{tabular}
  \bigskip
\end{minipage}
where the last term is a sum over the index $i$ that enumerates the
extrinsic noise variables $\eta_i$ (Einstein notation).
The $\Gamma_{r_i r_j}$ symbol has been evaluated according to
Eq.~\eqref{eq:delta1}.
We obtain the first order result in the variances $\epsilon_i^2$ of the
stochastic variables $\mu_i(t)$ from which we constructed the lognormal
variables (Eq.~\ref{eq:def_eta})
\begin{align}
  \average{\mathbf x^s} &= \mb\phi^s(\vect{0})
  + \frac{1}{2} \pfrac[\unskip^2]{\eta_i^2} \left.\mb\phi^s(\mb\eta)\right|_{\mb\eta = \vect{0}} \epsilon_i^2
  + \mathcal{O}(\epsilon_i^4) .
\end{align}
The contribution $\mathbf V(\mb\xi)$ in Eq.~\eqref{eq:result_Vxi} to the total
variance is formally equivalent and we obtain
\begin{align}
  \mathbf V(\mb\xi) &= \bC(\vect{0})
  + \frac{1}{2} \pfrac[\unskip^2]{\eta_i^2} \left.\bC(\mb\eta)\right|_{\mb\eta = \vect{0}} \epsilon_i^2
  + \mathcal{O}(\epsilon_i^4) .
\end{align}
As opposed to Eq.~\eqref{eq:taylor_C_s}, $\pfrac{\eta_i}
\left.\bC(\mb\eta)\right|_{\mb\eta = \vect{0}}$ is a Taylor
coefficient of $\bC(\mb\eta) \equiv \bC(\mb\eta, \mb\eta)$ from
Eq.~\eqref{eq:def_C} at equal times.

For the purely extrinsic contribution to the variance in Eq.~\eqref{eq:result_Vphis}
\begin{align*}
  \mathbf V(\mb\phi^s) =
  \sum\limits_{u = 1}^{\infty}
  \sum\limits_{n = 2}^{2 u}
  \sum\limits_{|\bq| = n}
  \hspace{-1.2em}\sum\limits_{\hspace{1.2em}\# \br^1 = q_1}
  \hspace{-1.2em}\sum\limits_{\hspace{1.2em}\# \br^2 = q_2}
  \mb\phi^s_{(\br^1)} \mb\phi^{s\,T}_{(\br^2)}\;
  \hspace{-.7em}\sum_{\hspace{.7em}|\bd^n| = u}\raisebox{3pt}{\hspace{-.9em}$''$}\hspace{0.5em}
  \,\prod_{
    \substack{ i, j = 1 \\ i < j }
  }^{n}
  \frac{1}{d_{ij}!}
  \Bigl(\Gamma_{f^2_i f^2_j}\Bigr)^{d_{ij}}
\end{align*}
the sum starts with $u=1$ and with the double prime only terms with derivatives of
both $\mb\phi^s$ and ${\mb\phi^s}^T$ contribute,
\\\begin{minipage}{.92\textwidth}
  \bigskip
  \begin{tabular}{p{1.5cm}p{2.0cm}p{2.0cm}p{2.0cm}l}
    $u=1$ & $\bq=(2,0)$ & $\br^1=(i,j)$ & $\br^2=()$    & 0 \\
	  & $\bq=(1,1)$ & $\br^1=(i)$   & $\br^2=(j)$   & \(
	  \mb\phi^s_{(i)} {\mb\phi^s_{(j)}}\!^T \,\Gamma_{i j}
	  = \left(\pfrac{\eta_i} \left.\mb\phi^s(\mb\eta)\right|_{\mb\eta = \vect{0}}\right)
	    \left(\pfrac{\eta_j} \left.\mb\phi^s(\mb\eta)\right|_{\mb\eta = \vect{0}}\right)^T \delta_{ij} \epsilon_i^2
	  \)\\
	  & $\bq=(0,2)$ & $\br^1=()$    & $\br^2=(i,j)$ & 0 \,.
  \end{tabular}
  \bigskip
\end{minipage}\\
The $\Gamma_{f^2_i f^2_j}$ symbol has been evaluated with
Eq.~\eqref{eq:delta1} and \eqref{eq:delta2}.
The first order result in the variances $\epsilon_i^2$ is
\begin{align}
  \mathbf V(\mb\phi^s) &= \left(\pfrac{\eta_i} \left.\mb\phi^s(\mb\eta)\right|_{\mb\eta = \vect{0}}\right)
  \left(\pfrac{\eta_i} \left.\mb\phi^s(\mb\eta)\right|_{\mb\eta = \vect{0}}\right)^T \epsilon_i^2
  + \mathcal{O}(\epsilon_i^4) .
\end{align}
The total variance is
\(
  \mathbf V\left(\mathbf x^s\right)
  = \mathbf V\left(\mb\phi^s\right)
  + \frac{1}{\Omega} \mathbf V\left(\mb\xi\right)
\)
according to Eq.~\eqref{eq:tscale_Vstx}.

In comparison to $\mathbf V(\mb\phi^s)$, the purely extrinsic contribution to
the spectrum matrix $\bP_e(\omega)$ in Eq.~\eqref{eq:result_Pe} needs evaluation
of the additional factor $\frac{\pi^{-1} \Theta}{\omega^2 + \Theta^2}$ with
$\bd^2 = (d_{12}) = (1)$, $\br^1=(i)$ and $n = 2$.
Eq.~\eqref{eq:integrals_Pe_final} then gives $\Theta(\bd^2, \br^1) = K_{i}$ and
\begin{align}
  \bP_e(\omega) &= \left(\pfrac{\eta_i} \left.\mb\phi^s(\mb\eta)\right|_{\mb\eta = \vect{0}}\right)
  \left(\pfrac{\eta_i} \left.\mb\phi^s(\mb\eta)\right|_{\mb\eta = \vect{0}}\right)^T
  \frac{K_{i} \epsilon_i^2}{\pi \left(\omega^2 + K_{i}^2\right)}
  + \mathcal{O}(\epsilon_i^4) .
\end{align}

For the second spectrum matrix $\bP_e(\omega)$
we need to evaluate Eq.~\eqref{eq:result_R},
\begin{align}
  \nonumber
  \bR(\omega)
    &=
    \sum\limits_{u=0}^{\infty}
    \sum\limits_{n=0}^{2u}
    \sum\limits_{a=0}^{n}
    \hspace{-1.4em}\sum\limits_{\hspace{1.4em}|\bq| = n - a}
    \hspace{-1.2em}\sum\limits_{\hspace{1.2em}\# \br^1 = q_1}
    \hspace{-0.5em}\ldots
    \hspace{-1.2em}\sum\limits_{\hspace{1.2em}\# \br^c = q_c}
    \hspace{-1.8em}\sum\limits_{\hspace{1.8em}\# \br^{c+1} = a}
    \hspace{-0.5em}\sum\limits_{\hspace{0.5em}\# \bs = a}
    \hspace{-.7em}\sum_{\hspace{.7em}|\bd^n| = u}\raisebox{3pt}{\hspace{-.9em}$'$}\hspace{0.5em}
  \\&\hspace{-1.4em}\times
  \nonumber
    \frac{
      1
    }{
      \Bigl(- \bA\!^0 + \theta(\bd^n, |\bq|, \br^{c+1}, \bs) + i \omega\Bigr)^{c + 1}
    }
    ~\bA_{(\br^1)} \ldots \bA_{(\br^c)} \bC_{(\br^{c+1},\bs)}
    \prod_{
      \substack{ i, j = 1 \\ i < j }
    }^{n}
    \frac{1}{d_{ij}!}
    \Bigl(\Gamma_{f^{c+1}_i f^{c+1}_j}\Bigr)^{d_{ij}} .
\end{align}
To elucidate the full complexity of the sum, we here evaluate some terms for
$u=2$.  We use an abbreviated notation for derivatives as
exemplarily defined by
\begin{align}
  \bA''_{ij} \equiv \frac{1}{2!} \pfrac{\eta_i} \pfrac{\eta_j} \left.\bA(\mb\eta)\right|_{\mb\eta=\vect{0}}
  \qquad\text{and}\qquad
  \bC_{i}^{\sigma} \equiv \pfrac{\eta_{i}^{\sigma}}
  \left.\bC(\mb\eta^1, \mb\eta^2)\right|_{\mb\eta^1 = \mb\eta^2 = \vect{0}}
  \,.
\end{align}
For $u = 2$, $n = 4$, $a = 1$ and $c = 2$ we obtain the terms
\begin{multline}
    \hspace{-0.5em}\sum\limits_{\hspace{0.5em} \substack{|\bq| = 3 \\ \# \bq = 2}}
    \hspace{-1.2em}\sum\limits_{\hspace{1.2em}\# \br^1 = q_1}
    \hspace{-1.2em}\sum\limits_{\hspace{1.2em}\# \br^2 = q_2}
    \sum\limits_{r^3_1}
    \sum\limits_{\substack{\hbox{\,}\vspace{0.2pt}\\\sigma_1}}
    \hspace{-.7em}\sum_{\hspace{.7em}|\bd^4| = 2}\raisebox{3pt}{\hspace{-.9em}$'$}\hspace{0.5em}
    \frac{1}{\Bigl(- \bA\!^0 + \theta(\bd^4, 3, \br^{3}, \bs) + i \omega\Bigr)^{3}}
    ~\bA_{(\br^1)} \bA_{(\br^2)} \bC_{(\br^{3},\bs)}
    \prod_{ \substack{ i, j = 1 \\ i < j } }^{4}
    \frac{1}{d_{ij}!} \Bigl(\Gamma_{f^{3}_i f^{3}_j}\Bigr)^{d_{ij}}
  \\
  = \left(
	\frac{\delta_{ij} \delta_{kl} \epsilon_i^2 \epsilon_k^2}{\Bigl(- \bA\!^0 + K_{kl} \delta_{2\sigma} + i \omega\Bigr)^{3}}
      + \frac{\delta_{ik} \delta_{jl} \epsilon_i^2 \epsilon_j^2}{\Bigl(- \bA\!^0 + K_{jl} \delta_{2\sigma} + i \omega\Bigr)^{3}}
      + \frac{\delta_{il} \delta_{jk} \epsilon_i^2 \epsilon_l^2}{\Bigl(- \bA\!^0 + K_{il} \delta_{2\sigma} + i \omega\Bigr)^{3}}
    \right)
    \left( \bA'_{i} \bA''_{jk} + \bA''_{ij} \bA'_{k} \right)\bC_{l}^{\sigma} .
\end{multline}
The sum over $|\bq| = 3$ was evaluated by writing $c=2$ symbols $\bA$ with all
possibilities to assign at least one of a total of three indices to each of them.
The Einstein notation for the extrinsic noise components
$i, j, k, l \in \{1,\dots,M\}$ of $\mb\eta$ and times $t_\sigma$, $\sigma \in
\{1, 2\}$ accounts for all other sums except the primed sum over $|\bd^4| = 2$.
The latter involves 3 tuples $(d_{12}, d_{13}, d_{23}, d_{14}, d_{24}, d_{34})$,
namely all components zero but $d_{12} = d_{34} = 1$, $d_{13} = d_{24} = 1$ or
$d_{14} = d_{23} = 1$.
The covariances $\Gamma_{f^3_i f^3_j}$ and the $\theta$ function evaluate
according to
Eq.~\eqref{eq:delta1}, \eqref{eq:delta2} and \eqref{eq:integrals_Pi_final}.

With $n = 3$ instead, $\bq$ needs to be $(1, 1)$ and we sum over
$|\bd^3| = 2$ and obtain the three terms
\begin{align}
  \left(
	\frac{\delta_{ij} \delta_{ik} \epsilon_i^4}{\left(- \bA\!^0 + K_{ik} \delta_{2\sigma} + i \omega\right)^{3}}
      + \frac{\delta_{ij} \delta_{jk} \epsilon_i^4}{\left(- \bA\!^0 + K_{jk} \delta_{2\sigma} + i \omega\right)^{3}}
      + \frac{\delta_{ik} \delta_{jk} \epsilon_i^4}{\left(- \bA\!^0 + (K_{ik} + K_{jk}) \delta_{2\sigma} + i \omega\right)^{3}}
    \right)
    \bA'_{i} \bA'_{j} \bC_{k}^{\sigma} .
\end{align}

Finally, we evaluate the terms for $u = n = 2$ and $a = 0$
for which $|\bd^2| = 2$ only allows $d_{12} = 2$
and with $a = 0$ no inverse correlation times $K_i$ are involved
but $\bq$ may be both $(2)$ with $c=1$ or $(1, 1)$ with $c=2$
which gives
\begin{align}
    \frac{\frac{1}{2!}\delta_{ij}^2 \epsilon_i^4}{\left(- \bA\!^0 + i \omega\right)^{2}} \bA''_{ij} \bC_0
  + \frac{\frac{1}{2!}\delta_{ij}^2 \epsilon_i^4}{\left(- \bA\!^0 + i \omega\right)^{3}} \bA'_i \bA'_j \bC_0
  = \frac{\frac{1}{2}\epsilon_i^4}{\left(- \bA\!^0 + i \omega\right)^{2}} \left(
    \bA''_{ii}
    + \frac{1}{\left(- \bA\!^0 + i \omega\right)} \bA'_i \bA'_i
  \right) \bC_0 .
\end{align}
We note here that the matrix multiplication is non commutative so the
order of terms is important.
In general, using Einstein notation for fixed $u$ and $n$, one writes down all possible terms
$\bA \ldots \bA \bC$ with $n$ indices ($\bC_0$ without index is allowed but not
so for the Taylor coefficients of $\bA$) and then for each term evaluates the
remaining sum over $|\bd^n| = u$ in order to derive the factors containing the
$\theta$ function (here $c$ is the number of $\bA$-symbols) and covariances
$\Gamma_{ij}$.

The spectrum matrix
$\average{\bP_i(\omega)} = \frac{1}{2\pi}\left(\bR(\omega) + \bR(\omega)^{*T}\right)$
in zero'th order is the power spectrum of the Ornstein-Uhlenbeck process for
the concentrations $\mathbf{x}$ in the absence of extrinsic noise.  The
Ornstein-Uhlenbeck is obtained from Eq.~\eqref{eq:LNA} by setting
$\mb\eta$ to zero.
Explicitly, its power spectrum is given by the well known
result \cite{Gardiner1990}
\begin{align}
  \bR(\omega) &= \left(-\bA^{0} + i\omega\right)^{-1} \bC_0
  + \mathcal{O}\left(\epsilon_i^2\right)
  \\
  \Rightarrow
  \average{\bP_i(\omega)}
  &= \frac{1}{2 \pi} \left(- \bA^0 \!+ i \omega\right)^{-1} \bB
  \bB^T \left(- \bA^{0\,T} \!- i \omega\right)^{-1}
  + \mathcal{O}\left(\epsilon_i^2\right)
\end{align}
where we have used the Lyapunov equation (\ref{eq:def_C}) for the last equality.
According to Eq.~\eqref{eq:P_as_sum}, the total spectrum matrix is
$\bP(\omega) = \bP_e(\omega) + \frac{1}{\Omega} \average[]{\bP_{i}(\omega)}$.

\end{widetext}

\bibliography{Noise_In_Ex.bib}
\end{document}